\documentclass[preprint,preprintnumbers,aps,superscriptaddress,nofootinbib,tightenlines,floatfix]{revtex4-1}
\usepackage{amsmath,amssymb}
\usepackage{graphicx}
\usepackage{bm}
\usepackage{comment}
\usepackage{color}

\usepackage[T1]{fontenc}
\usepackage[latin9]{inputenc}
\usepackage{graphicx}
\usepackage{esint}
\usepackage{hyperref}

\newcount\hour \newcount\hourminute \newcount\minute 
\hour=\time \divide \hour by 60
\hourminute=\hour \multiply \hourminute by 60
\minute=\time \advance \minute by -\hourminute
\newcommand{\mydate}{\ \today \ - \number\hour :\number\minute}

\begin{document}

\preprint{\vbox{ 
\hbox{NT@UW-12-08}
}}

\title{\textbf{Restoration of Rotational Symmetry in the Continuum Limit
of Lattice Field Theories}}

\author{Zohreh Davoudi}
\email{davoudi@uw.edu}
\affiliation{Department of Physics,
  University of Washington, Box 351560, Seattle, WA 98195, USA}

\author{Martin J. Savage}
\email{savage@phys.washington.edu}
\affiliation{Department of Physics,
  University of Washington, Box 351560, Seattle, WA 98195, USA}

\date{\mydate}

\begin{abstract}
We explore 
how rotational invariance 
is systematically recovered 
from calculations on hyper-cubic lattices through the use 
of smeared lattice operators 
that smoothly evolve into continuum operators with definite angular momentum as the
lattice-spacing is reduced.
Perturbative calculations of the angular momentum violation 
associated with such operators at tree-level and at one-loop 
are presented
in  $\lambda\phi^{4}$ theory and QCD.
Contributions from these operators that violate rotational invariance occur at tree-level,
with coefficients that are suppressed by ${\cal O}\left(a^{2}\right)$ in the
continuum limit. 
Quantum loops do not modify this behavior in $\lambda\phi^{4}$, nor in QCD if
the gauge-fields are smeared over a comparable spatial region.
Consequently, the use of this type of operator should, in principle,  allow for Lattice QCD
calculations of the higher moments of the hadron structure functions.
\end{abstract}
\pacs{}
\maketitle
\tableofcontents
\vfill\eject


\section{Introduction
\label{sec:Intro}
}
\noindent
Lattice quantum chromodynamics (LQCD) is a numerical 
technique in which Euclidean space correlation
functions of QCD are calculated by a Monte-Carlo evaluation of the Euclidean
space path integral~\cite{Wilson}. 
The computational resources are now becoming available for LQCD to recover the
spectrum of mesons and baryons that have been observed in the laboratory, and
to make predictions of states with exotic quantum numbers that will be the
focus of future experimental efforts.  It is also providing 
precise determinations of the matrix element of weak operators that are
required to further constrain the mixing of the eigenstates of the weak
interaction, contained in the CKM matrix.
LQCD is allowing for a comprehensive 
description of 
the structure of nucleons, and more recently to
their interactions that are crucial to the field of nuclear physics.  
This marks the beginnings of a comprehensive program to determine
nuclear structure and dynamics directly from QCD.

Space-time is pixelated, or 
discretized, in LQCD calculations,
with the quarks residing on the lattice sites, and the
gluon fields residing on the links between lattice sites. 
The lattice spacing, $a$, the distance between adjacent lattice sites,
is required to be much smaller than the characteristic hadronic length
scale of the system under study. 
In principle, 
the effects of a finite lattice
spacing can be systematically removed by combining calculations of 
correlation functions at several lattice spacings with the low-energy
effective field theory (EFT) which explicitly includes  the 
discretization effects. 
This type of EFT is somewhat more complicated than its
continuum counterpart as it must reproduce matrix elements of the
Symanzik action constructed with higher dimension operators induced by
the discretization~\cite{SymanzikI,SymanzikII,Parisi}. While the action lacks
Lorentz invariance and rotational symmetry, it is constrained by
hyper-cubic symmetry.  As computers have finite memory and performance,
the lattice volumes are finite in all four space-time directions.
Generally, periodic boundary conditions (BC's) are imposed on the
fields in the space-directions (a three-dimensional torus), while
(anti) periodic BC's are imposed on the (quark) gauge-fields 
in the time-direction.
However, the conceptual and practical problems
arising from the explicit breaking of the space-time symmetries
of the continuum theory, down to those of a hyper-cubic lattice theory, 
remain a challenge in the continuum  extrapolation of classes of observables
calculated using LQCD.
One knows, however,
that as the lattice becomes finer, the full space-time symmetries
of the continuum are in fact approximately recovered
for observables involving wavelengths that are large  compared with the
scale of pixelation.\footnote{For some numerical illustrations of this recovery in $SU\left(2\right)$ lattice gauge theories, as well as the scalar $\phi^{4}$ theory, see Refs. \cite{LangI, LangII, LangIII}.} As a result,
a quantitative description of this restoration, as well as its implication
for calculation of lattice observables, is possible.

Efforts to reduce lattice artifacts and achieve a better behaved
theory in the continuum limit date back to early stages of development
of LQCD. Many that fall under the name of Symanzik improvement
include a systematic modification of the action in such a
way to eliminate $\mathcal{O}\left(a^{n}\right)$ terms from physical
quantities calculated with  LQCD at each order in perturbation 
theory~\cite{SymanzikI,SymanzikII,Parisi,WeiszI,WeiszII,Luscher,Curci,Hamber,Eguchi,Wetzel,Sheikholeslami}, or nonperturbatively.
However, as will be discussed, discretization effects are known to 
give rise to more subtle issues; the treatment of which turns out to be more
involved.
LQCD is commonly formulated on a hyper-cubic grid, as a result the
full (Euclidean) Lorentz symmetry group of the continuum is
reduced to the discrete symmetry group of a hyper-cube. As
the (hyper-) cubic group has only a finite number of irreducible representations
(irreps) compared to infinite number of irreps of the rotational group, a given irrep of the rotational group is not irreducible
under the (hyper-) cubic group. 
Consequently, one can not assign
a well-defined angular momentum 
to a lattice state, which is generally a linear combination
of infinitely many different angular momentum states (see for example Refs. 
\cite{JohnsonRC,Berg,Mandula}). 
In principle, one can identify
the angular momentum of a corresponding continuum state in a lattice
calculation
from the degeneracies in the spectrum of
states belonging to different irreps of the cubic group as the lattice
spacing is reduced (a review of baryon spectroscopy efforts is
given in Ref. \cite{Lin}, for some recent meson spectroscopy works see Refs.
 \cite{BurchI,Gattringer,Petry,BurchII}). 
However, 
as the density of degenerate states substantially increases with increasing
the angular momentum,
the identification of states with 
higher angular momentum becomes impossible with the current statistical precision. 
The other issue is that the cubic symmetry of the lattice
allows the renormalization mixing of interpolating operators with
lower dimensional ones.  
The induced coefficients of the lower-dimensional operators
scale as inverse powers of the
lattice spacing, and hence diverge 
as the lattice spacing goes to zero. 
Although renormalization mixing of operators
is familiar from the continuum quantum field theory, it happens more
frequently in LQCD calculations as the reduced symmetry of the hyper-cube
is now less restrictive in preventing operators from mixing.
To obtain useful results for, as an example, the matrix elements of operators
from LQCD calculations,
non-perturbative
subtraction of the power-divergences is required and generally introduces large statistical
uncertainties.

To overcome these obstacles, it has been recently proposed by Dudek,
{\it et al.}~\cite{DudekI,DudekII,Edwards} (and later applied to bbb system
by Meinel~\cite{Meinel}) that by means of a novel construction of
interpolating operators, the excited states of several mesons and
baryons can be identified to high precision. 
The essence of this
method is that if one uses a set of cubically invariant local
operators which have already been subduced \cite{Basak} from a rotationally
invariant local operator with a definite angular momentum, $J$, while at the same
time smearing the gauge and quark fields over
the hadronic scale~\cite{Allton,Morningstar,Peardon}, the constructed
operator has maximum overlap onto a continuum state with angular momentum
$J$ if the lattice spacing is sufficiently small. 
The subduction is assumed to be responsible for retaining ``memory'' of the 
underlying angular momentum of the continuum operator, while 
the smearing is assumed to suppress mixing with operators of different angular momentum
by filtering contributions from ultra-violet (UV) modes.
In another approach, states with  higher angular momentum  
in the glueball spectra
of $2+1$ dimensional $SU\left(2\right)$ gauge theories~\cite{Meyer,JohnsonRW}
are isolated by using glueball interpolating operators that are 
linear
combinations of Wilson loops which are rotated by arbitrary angles
in order to project out a particular angular momentum $J$ in the continuum.
In addition, 
the links are smeared, or blocked, in order to be smooth over  physical
length scales  rather than just in the UV~\cite{Teper}.
So by monitoring the angular content of the glueball wavefunction
in the continuum limit with a probe with definite $J$, 
the $0^{-}/4^{-}$ puzzle in the glueball
spectroscopy has been tackled.
The prominent feature of these works
is that the recovery of rotational symmetry for sufficiently small
lattice spacings is qualitatively emergent from their numerical results.

The same issue occurs in LQCD calculations of higher moments
of hadron structure functions, the extraction
of which requires the matrix elements of local operators
between hadronic states.
Although 
Lorentz invariance forbids
twist-2 operators with different $J$
from mixing in the continuum, generally 
they can mix in LQCD calculations with power-divergent mixing 
coefficients~\cite{Capitani,Beccarini}.
The power-divergent  mixing problem associated with the  lower
moments can be avoided  by several means as described, for example, in Refs. ~\cite{Beccarini,GockelerI,GockelerII,GockelerIII,GockelerIV,GockelerVI,GockelerV,MartinelliI,MartinelliII,MartinelliIII,GockelerVI}. 
In addition to these approaches,
two methods~\cite{Dawson,Detmold} have been suggested
that highlight the idea of approaching the continuum properties
of the hadronic matrix elements by suppressing the contributions from the UV, 
and in that sense resemble the idea of operator smearing in the proposals described above. 
In LQCD calculations of non-leptonic K-decay, 
Dawson {\it et al.}~\cite{Dawson}
suggested that point-splitting the hadronic currents by a distance
larger than the lattice spacing, but smaller than the
QCD scale, results in an operator product expansion (OPE) of
the currents with the coefficients of lower dimensional operators
scaling with inverse powers of the point-splitting distance,
as opposed to the inverse lattice spacing.
This  considerably reduces  the numerical issues
introduced by  the operator mixing. 
In a different, but still physically equivalent approach, Detmold and Lin~\cite{Detmold}
showed that in the LQCD calculation of matrix elements of the Compton
scattering tensor, 
the introduction of a fictitious, non-dynamical, heavy
quark coupled to physical light quarks 
removes the power divergences of the mixing
coefficients.
This technique enables the extraction of matrix elements
of higher spin twist-2 operators with  a simple
renormalization procedure. 
The essence of this method is that the
heavy quark propagator acts as a smearing function in the momentum-space,
suppressing contributions from the high energy modes, provided
that its mass is much smaller than the inverse lattice spacing.

Encouraged by the results of the 
numerical non-perturbative investigation
of  Refs.~\cite{DudekI,DudekII,Edwards} 
and  Refs.~\cite{Meyer,JohnsonRW},
as well as the results of
Refs.~\cite{Dawson,Detmold}, 
we aim to quantify the recovery of rotational symmetry 
with analytical, perturbative calculations  in
$\lambda\phi^4$ and  QCD. 
In order to
achieve this goal, we first define a composite operator on the lattice
which has a well-defined angular momentum in the continuum limit and
is smeared over a finite physical region, and show how the non-continuum
contributions to the multipole expansion of the operator scales as
the lattice spacing is reduced toward the continuum. 
Tree-level contributions to matrix elements that violate
rotational symmetry,
either by the lattice operator
matching onto continuum operators with the ``wrong'' angular
momentum,
or matching onto continuum operators that explicitly 
violate rotational symmetry,
scale as $\mathcal{O}\left(a^{2}\right)$
as $a\rightarrow 0$. 
This includes the (naively) power divergent contributions from
lower-dimension operators.
In order to make definitive statements about the size of 
violations to rotational symmetry, it must be ensured that the 
tree-level scalings are not ruined by quantum
fluctuations. 
This is  demonstrated by a perturbative calculation of the two-point function in 
$\lambda\phi^{4}$ scalar field theory with an insertion of such an operator.
It is confirmed that quantum
corrections at any order in perturbation theory do not alter the observed
classical scalings of non-continuum contributions. This result is
comparable with finite size scaling results of the leading irrelevant
operator that breaks rotational invariance in three dimensional $O\left(N\right)$
models given in Refs.~\cite{CampostriniI,CampostriniII}. 
The
critical exponent $\rho$ introduced there, has a realization in terms
of small-$a$ scaling of the leading rotational invariance violating
terms in this calculation. Its value is shown to be consistent
with the results presented here.

After gaining experience with this operator in scalar field theory, 
the generalization to gauge theories is straightforward.
Special attention must be paid to the gauge links that appear in the
definition of gauge-invariant operator(s) that are the analogue of those
considered in the scalar field theory.
Also, it is well known that the perturbative expansion of operators used in
LQCD  are not well-behaved due to the presence of tadpole
diagrams~\cite{Lepage}. 
Naively, tadpoles make enhanced contributions to the matrix elements of the
operators we consider, and that tadpole improvement of the gauge links  and
smearing of the gluon fields 
are
crucial to the suppression of violations of rotational symmetry. 
After discussing the continuum behavior of the QCD operator(s), and their potential mixings, which violate rotational invariance at ${\cal O}(a^2)$,
we determine the renormalization of the operator(s) on the lattice
at one-loop order. 
The leading rotational invariance violating contributions to the renormalized
lattice 
operator are suppressed by ${\cal O}(\alpha_{s} a^2)$, 
(where $\alpha_{s}=g_s^2/(4\pi)$ and $g_s$ is the strong coupling constant)
provided that the gauge
fields 
are also smeared over a physical region similar to the matter fields. 
This means that the leading rotational invariance violating operators
introduced by the quantum loops make subleading contributions 
compared to tree-level, ${\cal O} (a^2)$.
The loop contributions that scale as ${\cal O}(\alpha_{s} a)$ 
do not violate rotational symmetry, and hence 
are absorbed into the operator Z-factor.

\section{Operators in Scalar Field Theory
\label{sec:Classical}}
\noindent 
\begin{figure}
\begin{centering}
\includegraphics[scale=0.50]{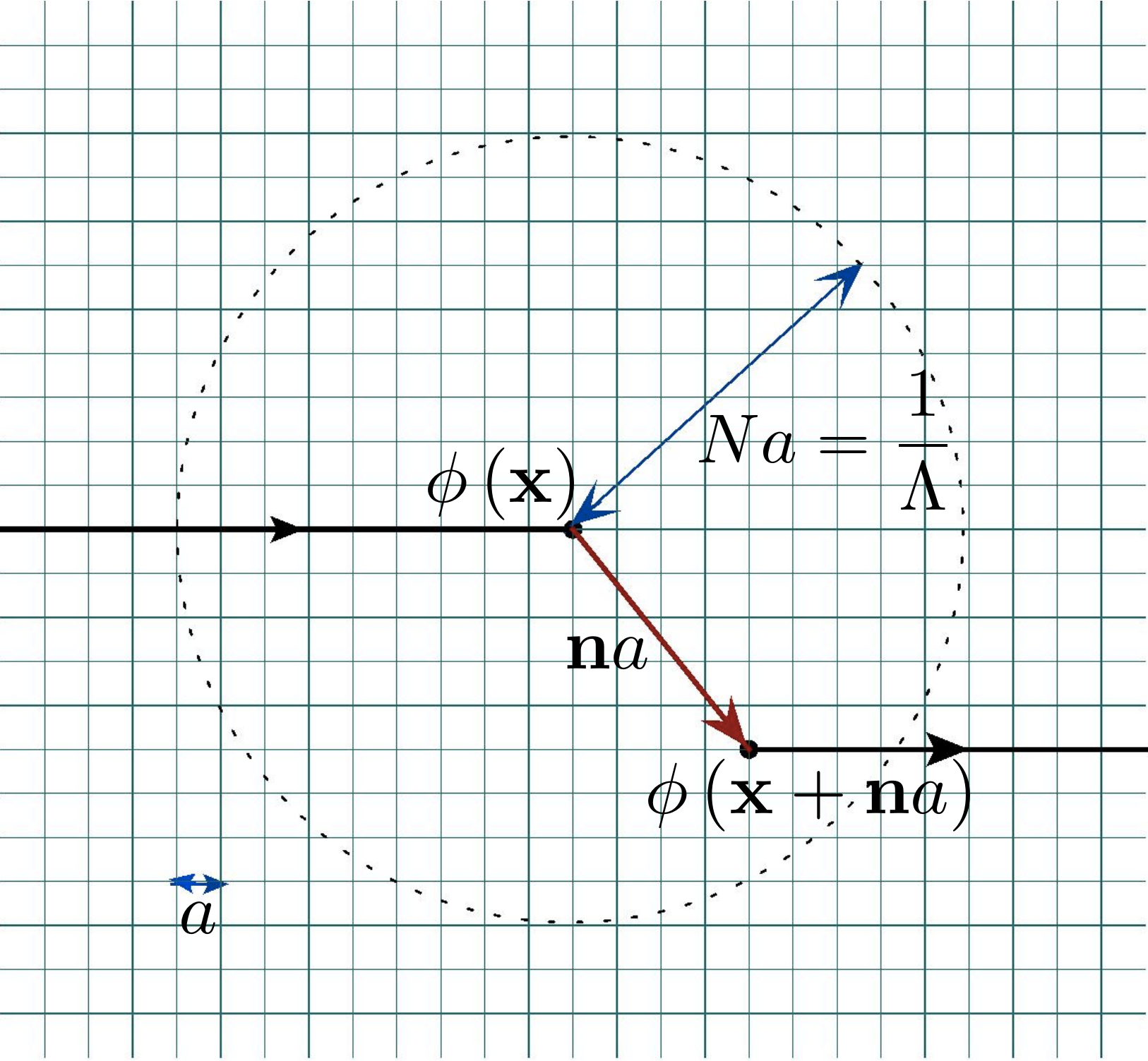}
\par\end{centering}
\caption{{\small 
A contribution to the lattice operator defined in 
Eq.~(\ref{eq:1}), with $\left|\mathbf{n}\right|\leq N$. 
All the points inside the three-dimensional spherical shell
$\left|\mathbf{n}a\right|=Na$ are included in the operator. 
The two length scales defining the operator, 
the lattice spacing, $a$, 
and the operator size, $Na=1/\Lambda$, 
are shown.}}
\label{fig:operator}
\end{figure}
The goal is to construct a bilinear operator of the scalar fields
on a cubic lattice which has certain properties. First of all, as
it was discussed earlier, it has to be smeared over a finite region
of space. 
This physical region should be large compared to the lattice
spacing, 
and, for our purposes,  small compared to typical length scale of
the system  to allow for a perturbative analysis.
The spatial extent of the operator can be identified with its renormalization scale.
Secondly, 
it is required to transform as a spherical tensor with well-defined angular
momentum in the continuum limit.
An operator that satisfies these conditions is~\footnote{
This corresponds to one particular choice of radial structure of the operator.
However, the results of the calculations  and the physics conclusions presented
in this work do not change qualititively when other smooth radial structures are
employed, such as a Gaussian or exponential.
} 
\begin{equation}
\hat{\theta}_{L,M}\left(\mathbf{x};a,N\right)
\ =\ 
\frac{3}{4\pi
  N^{3}}
\sum_{\mathbf{n}}^{\left|\mathbf{n}\right|\leq N}\phi\left(\mathbf{x}\right)
\phi\left(\mathbf{x}+\mathbf{n}a\right)
\ Y_{L,M}\left(\hat{\mathbf{n}}\right)
\ \ \ ,
\label{eq:1}
\end{equation}
where $\mathbf{n}$ denotes a triplet of integers, and  
it is normalized
by the spatial volume of the region over which it is distributed.
$\phi({\bf x})$ is the scaler field operator, 
$N$ is the maximum number of lattice sites in the radial direction, and 
$Y_{LM}\left(\hat{\mathbf{n}}\right)$ is a spherical harmonic evaluated at the
angles defined by the unit vector in the direction of  $\mathbf{n}$, 
$\hat{\mathbf{n}}$, as shown in fig.~\ref{fig:operator}.
This operator can also be written in a multipole expansion about its center as
\begin{equation}
\hat{\theta}_{L,M}\left(\mathbf{x};a,N\right)=\frac{3}{4\pi
  N^{3}}
\sum_{\mathbf{n}}^{\left|\mathbf{n}\right|\leq N}\sum_{k}\frac{1}{k!}\ 
\phi\left(\mathbf{x}\right)\left(a\mathbf{n}\cdot\mathbf{\nabla}\right)^{k}
\phi\left(\mathbf{x}\right)
\ Y_{L,M}\left(\hat{\mathbf{n}}\right)
\ \ \ ,
\label{eq:2}
\end{equation}
where the gradient operator acts on the ${\bf x}$ variable,
$\mathbf{\nabla}\equiv \mathbf{\nabla}_{\bf x}$.

Although the operator $\hat{\theta}_{L,M}\left(\mathbf{x};a,N\right)$
is labeled  by its angular momentum in the continuum limit, from the right hand side of
eq.~(\ref{eq:2}), it is clear that it is a linear combination
of an infinite number of operators with angular momentum compatible with its parity. 
To be more specific, consider the $M=0$
component of the operator expanded in a derivative  operator basis,
\begin{equation}
\hat{\theta}_{L,0}\left(\mathbf{x};a,N\right)=
\sum_{L^{\prime},d}\frac{C_{L0;L^{\prime}0}^{\left(d\right)}
\left(N\right)}{\Lambda^{d}}\mathcal{O}_{z^{L^{\prime}}}^{\left(d\right)}
\left(\mathbf{x};a\right)
\ \ \ ,
\label{eq:3}
\end{equation}
where $\mathcal{O}_{z^{L^{\prime}}}^{\left(d\right)}\left(\mathbf{x};a\right)$
are defined in Appendix~\ref{app:operators}.
The operator subscript denotes  that there are $L^{\prime}$ 
free indices in the derivative operator, while $d$ denotes the total number of derivatives. 
As is discussed in the
Appendix~\ref{app:operators}, 
there are operators in this basis which are not rotationally
invariant but only cubically invariant. 
$C_{L0;L^{\prime}0}^{\left(d\right)}\left(N\right)$
are coefficients of each operator in the expansion whose values are
determined by matching eq.(\ref{eq:2}) with eq.~(\ref{eq:3}). 
Finally $\Lambda=1/(Na)$ is the momentum-scale of the smeared
operator which is kept fixed as the lattice spacing is varied. 
Therefore, as the lattice spacing decreases,
more point shells (shells of integer triplets) 
are included in the sum in eq.~(\ref{eq:2}).
The convergence of this derivative expansion is guaranteed
as the scale $\Lambda$ is set to be much larger than the typical
momentum encountered by the operator.

\subsection{Classical Scalar Field Theory}
\label{sec:Clasphi}

In order for the operator to recover its continuum limit as the lattice
spacing vanishes, 
the coefficients $C_{L0;L^{\prime}0}^{\left(d\right)}$
should have certain properties. 
First of all, those associated with
the operators with $L\neq L^{\prime}$ as well as the rotational invariance
violating operators, should vanish as $a\rightarrow0$.
Also the
coefficients of rotational invariant operators with $L=L^{\prime}$
should reach a finite value in this limit. 
These properties will be
shown to be the case in a formal way shortly, but in order to get
a general idea of the classical scaling of the operators and the size
of mixing coefficients, we first work out a particular example. 
Consider
the operator $ $$\hat{\theta}_{3,0}\left(\mathbf{x};a,N\right)$
expanded out up to five derivative operators,
\begin{equation}
\begin{array}{c}
\hat{\theta}_{3,0}\left(\mathbf{x};a,N\right)
=
\frac{C_{30;10}^{\left(1\right)}\left(N\right)}{\Lambda}\mathcal{O}_{z}^{\left(1\right)}
\left(\mathbf{x};a\right)
\ +\ 
\frac{C_{30;10}^{\left(3\right)}\left(N\right)}{\Lambda^{3}}\mathcal{O}_{z}^{\left(3\right)}
\left(\mathbf{x};a\right)
\ +\ 
\frac{C_{30;10}^{\left(5\right)}\left(N\right)}{\Lambda^{5}}\mathcal{O}_{z}^{\left(5\right)}
\left(\mathbf{x};a\right)
\ + \  \\
\\
\frac{C_{30;10}^{\left(5;RV\right)}\left(N\right)}{\Lambda^{5}}\mathcal{O}_{z}^{\left(5;RV\right)}
\left(\mathbf{x};a\right)
\ +\ 
\frac{C_{30;30}^{\left(3\right)}\left(N\right)}{\Lambda^{3}}\mathcal{O}_{zzz}^{\left(3\right)}
\left(\mathbf{x};a\right)
\ +\ 
\frac{C_{30;30}^{\left(5\right)}\left(N\right)}{\Lambda^{5}}\mathcal{O}_{zzz}^{\left(5\right)}
\left(\mathbf{x};a\right)
\ +\ \\
\\
\frac{C_{30;50}^{\left(5\right)}\left(N\right)}{\Lambda^{5}}\mathcal{O}_{zzzzz}^{\left(5\right)}
\left(\mathbf{x};a\right)
\ +\ 
\mathcal{O}\left(\frac{\nabla_{z}^{7}}{\Lambda^{7}}\right)
\ \ \ ,
\end{array}
\label{eq:4}
\end{equation}
where the superscript RV denotes the rotational invariance violating operator and its corresponding coefficient in the above expansion.
\begin{figure}[!ht]
\begin{centering}
\includegraphics[scale=1]{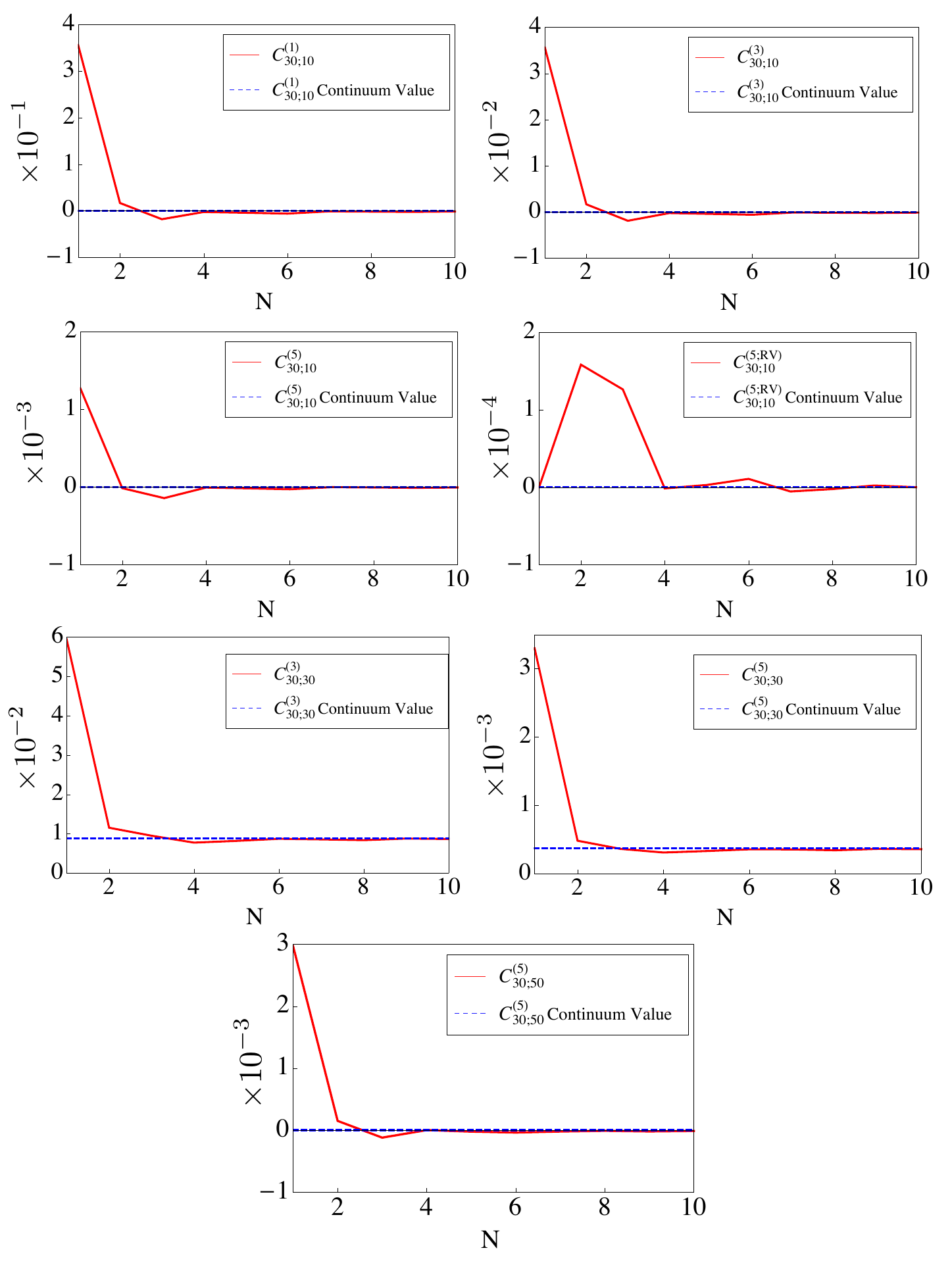}
\par\end{centering}
\caption{{\small 
The tree-level values of the 
coefficients $C^{(d)}_{30;L^\prime 0}$  appearing in
eq.~(\protect\ref{eq:4})
as a function of the largest $n$-shell included in the summation in eq.~(\protect\ref{eq:1}).
}}
\label{fig:TheCs}
\end{figure}
\begin{figure}[!ht]
\begin{centering}
\includegraphics[scale=0.5]{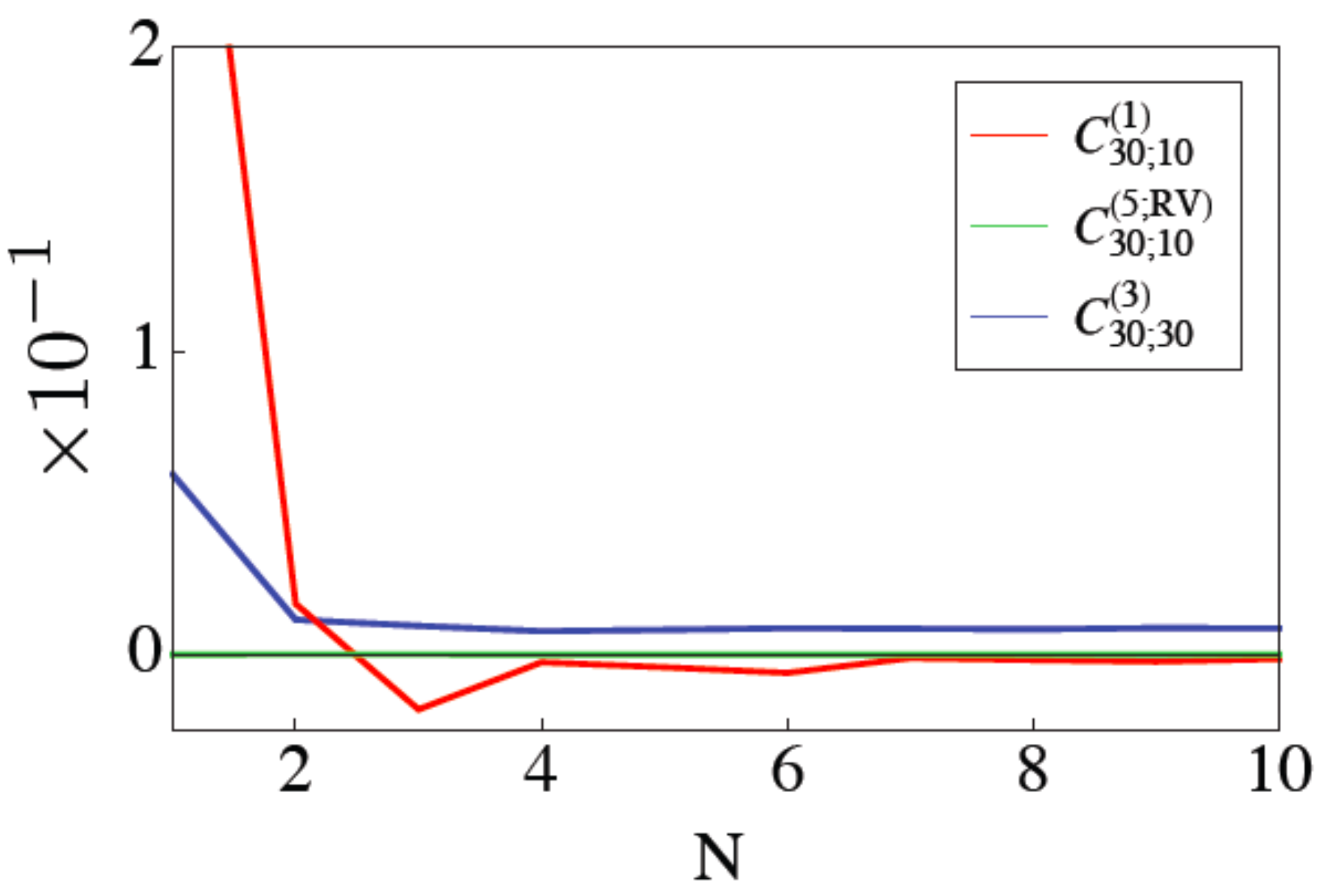}
\par\end{centering}
\caption{{\small 
A comparison between the tree-level coefficients $C^{(d)}_{30;L^\prime 0}$ to illustrate
the relative rates of convergence to the continuum limit.
}}
\label{fig:TheComp}
\end{figure}

The numerical values of the coefficients in eq.~(\ref{eq:4}), at the classical level,
as a
function of the maximum shell included in the sum in eq.~(\ref{eq:2})
are shown in fig.~\ref{fig:TheCs} and fig.~\ref{fig:TheComp}.
From these plots it is clear that while the coefficients
$C_{30;30}^{\left(3\right)}$ and $C_{30;30}^{\left(5\right)}$ reach
a finite value for large N, the coefficients of lower and higher angular
momentum operators,  as well as the rotational invariance violating
operator, approach zero. 
To find the values of the 
leading order (LO) 
coefficients in this limit, as well as to see how the non-leading
contributions scale with $N=1/(\Lambda a)$, one can apply the Poisson
re-summation formula to the right hand side of eq.~(\ref{eq:2}),
\begin{equation}
\hat{\theta}_{L,M}\left(\mathbf{x};a,N\right)
 =\ 
\frac{3}{4\pi N^{3}}\sum_{k}\frac{a^{k}}{k!}
\sum_{\mathbf{p}}\int d^{3}y\ 
\theta\left(N-y\right)\ 
e^{i2\pi\mathbf{p}\cdot\mathbf{y}}\ 
\phi\left(\mathbf{x}\right)\left(\mathbf{y}\cdot\mathbf{\nabla}\right)^{k}
\phi\left(\mathbf{x}\right)\ 
Y_{L,M}\left(\hat{\mathbf{y}}\right)
\ \ \ ,
\label{eq:5}
\end{equation}
where $\mathbf{p}$ is another triplet of integers, and the
$\mathbf{p}$ summation is unbounded. 
The continuum values of the coefficients
obtained in the $N\rightarrow\infty$ limit, 
corresponding to the $\mathbf{p}=0$ term in eq.~(\ref{eq:5}),
are
\begin{eqnarray}
C_{30;30}^{\left(d\right)}
& = & 
{15\over 4}\ 
\sqrt{7\over\pi}\ 
{ d^2-1 \over (d+4)!}
 \qquad {\rm with}\qquad d=3,5,...
\ \ \ ,
\label{eq:6}
\end{eqnarray}
while the other coefficients 
in eq.~(\ref{eq:4})
vanish in this limit as expected.
The LO  corrections to these continuum values can be calculated
as following. 
The deviation of $C_{30;30}^{\left(3\right)}$ from
its continuum value can be found from 
\begin{eqnarray}
I_{30}
& \sim\ & 
\frac{3}{4\pi}\frac{\left(Na\right)^{3}}{3!}
\sum_{\mathbf{p}\neq 0}
\int_{0}^{1}dy\ y^{2}\ d\Omega_{\hat{y}}\ 
e^{i2\pi N\mathbf{p}\cdot\mathbf{y}}
\ \phi\left(\mathbf{x}\right)\ 
\ 
\left(\hat{\mathbf{y}}\cdot\mathbf{\nabla}\right)^{3}
\phi\left(\mathbf{x}\right)\ 
\ Y_{3,0}\left(\hat{\mathbf{y}}\right)
\ \ \ ,
\label{eq:7}
\end{eqnarray}
where $\mathbf{\nabla}=\nabla_{z}\hat{e}_{z}$ and the $y$-variable  in eq.~(\ref{eq:7})
is redefined to lie between $0$ and $1$,
and it is straightforward to show that
\begin{equation}
\delta C_{30;30}^{\left(3\right)}
\ =\ 
\frac{1}{N^{2}}
\ \frac{1}{32\pi^{2}}
\ \sqrt{\frac{7}{\pi}}\ 
\sum_{\mathbf{p}\neq0}
\frac{\cos\left(2\pi
    N\left|\mathbf{p}\right|\right)}{\left|\mathbf{p}\right|^{8}}
\left(-\frac{3}{2}\left|\mathbf{p}\right|^{6}+15\left|\mathbf{p}\right|^{2}p_z^4-\frac{25}{2}p_{z}^{6}\right)
\ \ \ .
\label{eq:8}
\end{equation}
It is interesting to note that, after trading $N$ for $1/(a \Lambda)$,
the finite lattice spacing corrections are not monotonic in $a$,
but exhibit oscillatory behavior, which is clearly evident in fig.~\ref{fig:TheCs}.

The  deviation of $C_{30;10}^{\left(1\right)}$ from its continuum value of zero
follows similarly, and is found to scale as $\sim 1/N^2$, 
\begin{equation}
\delta C_{30;10}^{\left(1\right)}
\ =\ \frac{1}{N^{2}}
\ \frac{3}{16\pi^{2}}\sqrt{\frac{7}{\pi}}
\ \sum_{\mathbf{p}\neq0}
\ \frac{\cos\left(2\pi
    N\left|\mathbf{p}\right|\right)}{\left|\mathbf{p}\right|^{6}}
\ \left(\left|\mathbf{p}\right|^{4}-5p_{z}^{4}\right)
\ \ \ \ .
\label{eq:9}
\end{equation}
As in the case of the operator that conserves angular momentum in the continuum
limit,
the sub-leading correction (and in this case the first non-zero contribution) 
to the coefficient is suppressed by $1/N^{2}$. 
This can be shown to be the case for all the sub-leading
contributions to the coefficients $C_{LM;L^{\prime}M^{\prime}}^{\left(d\right)}$ as follows.
As is evident from eq.~(\ref{eq:5}), the integrals
that are required
in calculating deviations from the continuum
values have the general form
\begin{equation}
I^{i_1...i_k}
\ \sim\ 
\frac{3}{4\pi}\frac{\left(Na\right)^{k}}{k!}
\ \sum_{\mathbf{p}\neq0}
\ \int_{0}^{1}dy\ y^{2+k}\ \int d\Omega_{\hat{y}}
\ e^{i2\pi N\mathbf{p}\cdot \mathbf{y}}\ \hat{y}^{i_{1}}\
\hat{y}^{i_{2}}...\hat{y}^{i_{k}}
\ Y_{LM}\left(\Omega_{\hat{y}}\right)
\ \ \ ,
\label{eq:10}
\end{equation}
which can be written as
\begin{eqnarray}
I^{i_1...i_k}
&  \sim & 
\frac{3}{4\pi}
\ \frac{\left(Na\right)^{k}}{k!}\frac{1}{\left(i2\pi N\right)^{k}}
\ \sum_{\mathbf{p}\neq0}\frac{\partial}{\partial
  p_{i_{1}}}...\frac{\partial}{\partial p_{i_{k}}}
\ \int_{0}^{1}dy\ y^{2+k}\ \int d\Omega_{\hat{y}}
\ e^{i2\pi N\mathbf{p}\cdot\mathbf{y}}
\ Y_{LM}\left(\Omega_{\hat{y}}\right)
\nonumber\\
& \sim & 
\frac{3}{4\pi}\frac{\left(Na\right)^{k}}{k!}\ 
\frac{4\pi i^{L}}{\left(i2\pi
    N\right)^{k}}\sum_{\mathbf{p}\neq0}
\ \frac{\partial}{\partial
  p_{i_{1}}}...\frac{\partial}{\partial
  p_{i_{k}}}
\ Y_{LM}\left(\Omega_{\hat{p}}\right)
\ \int_{0}^{1}dy\ y^{2+k}\ j_{L}\left(2\pi N\left|\mathbf{p}\right|y\right)
\ \ .
\label{eq:11}
\end{eqnarray}
The y integration over the Bessel function gives rise to
either 
$-\frac{\cos\left(2\pi N\left|\mathbf{p}\right|\right)}{\left(2\pi N\left|\mathbf{p}\right|\right)^{2}}$
or 
$-\frac{\sin\left(2\pi N\left|\mathbf{p}\right|\right)}{\left(2\pi N\left|\mathbf{p}\right|\right)^{2}}$,
up to higher orders in $1/N$, 
depending on whether $L$ is even or odd. 
Thus the LO contribution from
eq.~(\ref{eq:11}) in the  large $N$ limit 
is obtained by acting 
on the numerator
with the $p$ derivatives,  producing  $k$ powers
of $N$,  multiplying the $1/N^{2}$ from the denominator. 
Therefore, eq.~(\ref{eq:11}) scales as
\begin{equation}
I^{i_1...i_k}\ \sim\ 
\left(Na\right)^{k}\frac{1}{N^{k}}\frac{N^{k}}{N^{2}}
\ \sim\ {1\over\Lambda^k}\ {1\over N^2}
\ \ \ \ ,
\label{eq:12}
\end{equation}
and, 
in general,
the deviation of any coefficient from its  continuum value  
is suppressed by $1/N^{2}=\Lambda^{2}a^{2}$. 
This result implies that in calculating the matrix element of $L=3$ operator,
one has a derivative expansion of the form
\begin{eqnarray}
\Lambda^{3}\hat{\theta}_{3,0}\left(\mathbf{x};a,N\right)
& = & 
\alpha_{1}\ 
\frac{\Lambda^{2}}{N^{2}}\mathcal{O}_{z}^{\left(1\right)}\left(\mathbf{x};a\right)
\ +\ 
\alpha_{2}\ 
\frac{1}{N^{2}}\mathcal{O}_{z}^{\left(3\right)}\left(\mathbf{x};a\right)
\ +\ 
\alpha_{3}\ 
\frac{1}{\Lambda^{2}N^{2}}\mathcal{O}_{z}^{\left(5\right)}\left(\mathbf{x};a\right)
\nonumber\\
& + & 
\alpha_{4}\ 
\frac{1}{\Lambda^{2}N^{2}}\mathcal{O}_{z}^{\left(5;RV\right)}\left(\mathbf{x};a\right)
\ +\ 
\alpha_{5}\ 
\mathcal{O}_{zzz}^{\left(3\right)}\left(\mathbf{x};a\right)
 \ +\ 
\alpha_{6}\ 
\frac{1}{\Lambda^{2}}\mathcal{O}_{zzz}^{\left(5\right)}\left(\mathbf{x};a\right)
\nonumber\\
& + & 
\alpha_{7}\ 
\frac{1}{\Lambda^{2}N^{2}}\mathcal{O}_{zzzzz}^{\left(5\right)}\left(\mathbf{x};a\right)
 \ +\ 
\mathcal{O}\left({\nabla_{z}^{7}\over \Lambda^{4}}\right)
\ \ \ ,
\label{eq:13}
\end{eqnarray}
where the mixing with $L\neq3$ operators
(with coefficients $\alpha_{1,2,3,7,...}$),
as well as the
operator with broken rotational symmetry
(with coefficient $\alpha_4$), 
vanish in the  large $N$ limit, 
while
the coefficients of $L=3$ operators 
(with coefficients $\alpha_{5,6,...}$),
are fixed by the 
scale of the operator, $\Lambda$. 
It is clear that for $N=1$ and $\Lambda=1/a$, where no smearing
is performed, the problem with divergent coefficients of the lower
dimensional operators is obvious, as,  for example, the coefficient of
$\mathcal{O}_{z}^{\left(1\right)}\left(\mathbf{x};a\right)$ diverges
as $1/a^{2}$ as $a\rightarrow0$, as is well known.

The fact that all the sub-leading contributions to the classical operator
are suppressed at least by $1/N^{2}$ regardless of $L$ and $L^{\prime}$
can be understood as follows. 
In the classical limit, where the short
distance fluctuations of the operator are negligible, the operator
does not probe the distances of the order of lattice spacing when
$a\rightarrow 0$. 
The angular resolution of the operator is
dictated by the solid angle discretization of the physical region
over which the operator is smeared, and therefore is proportional to
$1/N^{2}$. 
The question to answer is whether the 
quantum fluctuations modify this general result.

Before proceeding with the quantum loop calculations, 
it is advantageous to transform the operator into 
momentum-space to simplify loop integrals.
This can be done
easily by noting that for zero momentum insertion,
the operator acting on the field with momentum
$\mathbf{k}$ is
\begin{equation}
\hat{\tilde{\theta}}_{LM}\left(\mathbf{k};a,N\right)
\ =\ 
\frac{3}{4\pi N^{3}}\ 
\sum_{\mathbf{n}}^{|\mathbf{n}|\leq N}
\ e^{i\mathbf{k}\cdot\mathbf{n}a}
\ Y_{LM}\left(\mathbf{n}\right)
\ \tilde{\phi}\left(\mathbf{k}\right)
\ \tilde{\phi}\left(-\mathbf{k}\right)
\ \ \ ,
\label{eq:14}
\end{equation}
which, after using the partial-wave expansion of
$e^{i\mathbf{k}\cdot\mathbf{n}a}$
and the exponential term resulting from the Poisson relation,
can be written as
\begin{eqnarray}
\hat{\tilde{\theta}}_{LM}\left(\mathbf{k};a,N\right)
& = &
6\sqrt{\pi}\ 
 \tilde{\phi}\left(\mathbf{k}\right)
 \tilde{\phi}\left(-\mathbf{k}\right)
\ \sum_{\mathbf{p}}\sum_{L_{1},M_{1},L_{2},M_{2}}i^{L_{1}+L_{2}}
\ \sqrt{\frac{\left(2L_{1}+1\right)\left(2L_{2}+1\right)}{2L+1}}
\nonumber\\
&&
\times
\ \left\langle L_{1}0;L_{2}0\left|L0\right.\right\rangle 
\ \left\langle L_{1}M_{1};L_{2}M_{2}\left|LM\right.\right\rangle 
Y_{L_{1}M_{1}}\left(\Omega_{\hat{k}}\right)
\ Y_{L_{2}M_{2}}\left(\Omega_{\hat{p}}\right)
\nonumber\\
&&
\times
\ \int_{0}^{1}dy\ y^{2}
\ j_{L_{1}}\left(aN\left|\mathbf{k}\right|y\right)
\ j_{L_{2}}\left(2\pi N\left|\mathbf{p}\right|y\right)
 \ \ \ .
\label{eq:15}
\end{eqnarray}
Although this form seems to be somewhat more complicated than in
position-space, 
it turns out that it is advantageous
to work in momentum-space when dealing with higher angular
momenta, as well as for $M\ne 0$.
Further, the dimensionless parameters 
$|{\bf k}|/\Lambda$ and $N$ that define the physics of such systems
are now explicit.
It is straightforward to show this form recovers 
the values of the
leading and sub-leading coefficients
given in eq.~(\ref{eq:8}) and eq.~(\ref{eq:9}),
and it is worth mentioning how they emerge from eq.~(\ref{eq:15}).
For a non-zero value of $|{\bf p}|$ and $N=\infty$, the spherical
Bessel function $j_{L_{2}}\left(2\pi N\left|\mathbf{p}\right|y\right)$ vanishes
for any value of $L_2$.  However, for large values of $N$ but $|{\bf p}|=0$ the
only non-zero contribution is from $L_2=0$, and thus $L_1=L$, leaving a
straightforward integration over a single spherical Bessel function
$j_{L}\left(aN\left|\mathbf{k}\right|y\right)$ to obtain the continuum limit
given in eq.~(\ref{eq:6}).
Extracting the subleading contributions and the violations of rotational
symmetry 
is somewhat more involved, and we provide an explicit
example in Appendix~\ref{app:RIviolation}.

\subsection{Quantum Corrections in $\lambda\phi^{4}$
\label{sec:Scalar}}
\noindent 
In order to determine the impact of quantum fluctuations on the matrix elements
of $\hat{\theta}_{L,M}$,
defined in eq.~(\ref{eq:1}), we consider loop contributions in
$\lambda\phi^{4}$ theory.
Beside its simplicity which enables us to develop tools in performing 
the analogous calculations in Lattice QCD, 
this theory corresponds to some interesting condensed matter systems.
For example, three dimensional O(N) models,  which describe important
critical phenomena in nature, have a corresponding $\lambda\phi^{4}$
field theory formulation. 
As pointed out in Refs.~\cite{CampostriniI,CampostriniII},
anisotropy in space either due to the symmetries of the physical system,
or due to an underlying lattice formulation, 
will result in the presence of irrelevant operators in the effective
Hamiltonian which are not rotationally invariant,
and 
introduce deviations of two-point functions
from their rotationally invariant scaling law near the fixed point. 
However,
as the rotationally invariant fixed point of the theory is approached,
the anisotropic deviations vanish like $1/\xi^{\rho}$ where $\xi^{2}$
is the second moment correlation length derived from the two-point function, 
and $\rho$ is a critical
exponent which is related to the critical effective dimension of the
leading irrelevant operator breaking rotational invariance. 
It has
been shown that in the large N approximation
of $O\left(N\right)$ models, $\rho\simeq2$ for cubic-like lattices. 
In the following, it
will be shown that, 
by inserting
$\hat{\theta}_{L,M}$ defined in
eq.~(\ref{eq:1}) into the two-point function,
the same scaling law emerges when approaching the rotational-invariant
continuum limit of $\lambda\phi^{4}$ theory.
\begin{figure}
\begin{centering}
\includegraphics[scale=0.15]{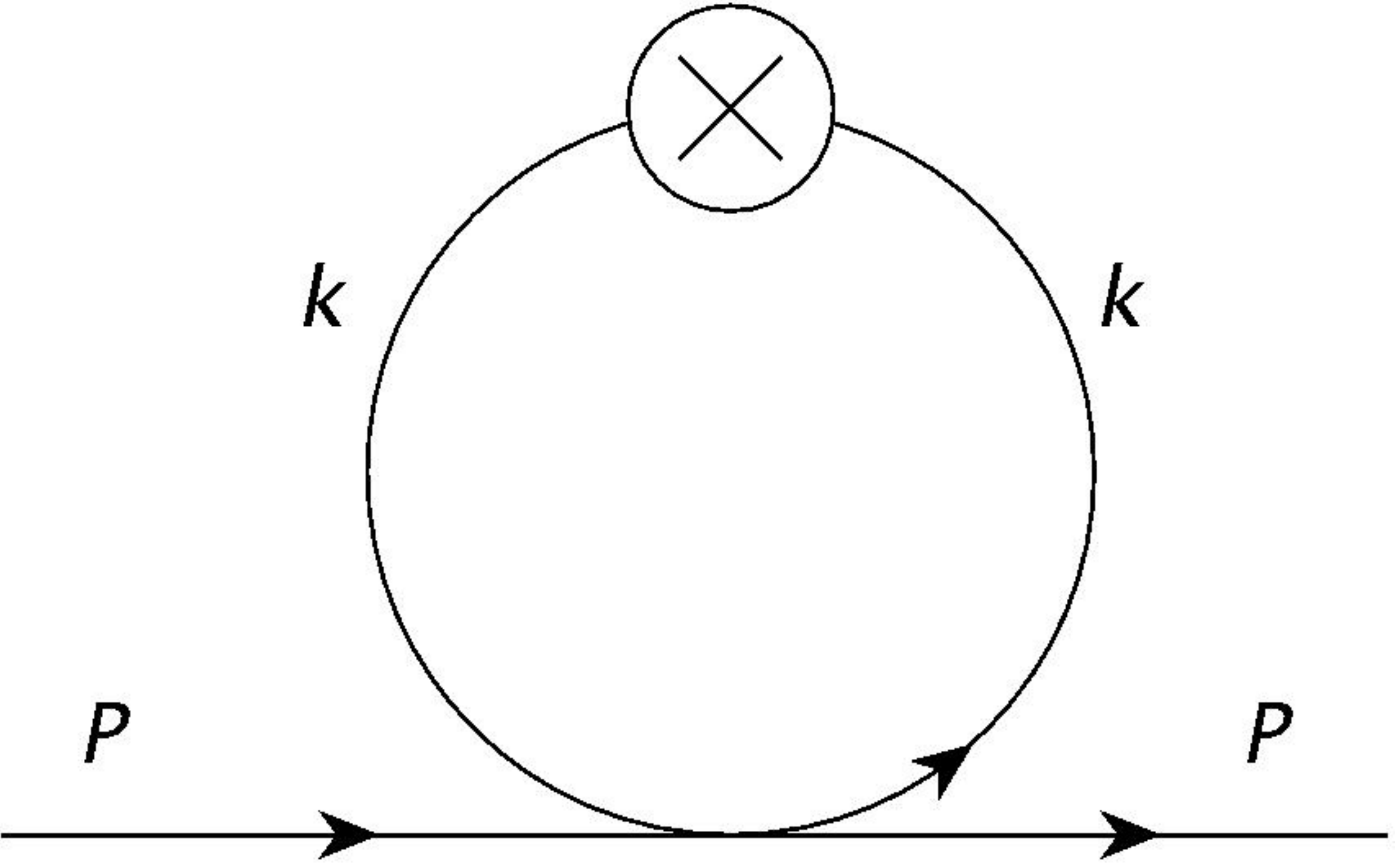}
\par\end{centering}

\caption{{\small One-loop correction to the two-point function with an insertion
of $\hat{\theta}_{L,M}$ in $\lambda\phi^{4}$}}
\label{fig:onelooplf4}
\end{figure}

At tree level, the contributions to the two-point function 
from an insertion of 
$\hat{\theta}_{L,M}$
at zero momentum transfer
has been already discussed in section \ref{sec:Clasphi}.
At one-loop, there is only one diagram with an insertion of $\hat{\theta}_{L,M}$
that contributes to the two-point function, as shown in
fig.~\ref{fig:onelooplf4}. 
This diagram introduces corrections only to the $L=0$ matrix element as there are
no free indices associated with the loop. 
The lattice integral associated with this one-loop diagram is 
\begin{equation}
J_{LM}
\ =\ 
\frac{3\lambda}{4\pi
  N^{3}}\sum_{\mathbf{n}}^{|\mathbf{n}|\leq N}
\ \int_{-\frac{\pi}{a}}^{\frac{\pi}{a}}\frac{d^{4}k}{\left(2\pi\right)^{4}}
\ \frac{e^{i\mathbf{k}\cdot\mathbf{n}a}}{\left(\hat{k}^{2}+m^{2}\right)^{2}}
\ Y_{LM}\left(\Omega_{\mathbf{n}}\right)
\ \ \ ,
\label{eq:17}
\end{equation}
where 
$\hat{k}^{2}= {4\over a^2}\sum\limits_{\mu} \sin^{2}\left({ k_{\mu}a\over
    2}\right)$, 
$\lambda$ is the coupling constant and $m$
is the $\phi$ mass.
The three-momentum integration can be
evaluated by noting that the region of integration
can be split into two parts: region I where $0\leq\left|\mathbf{k}\right|\leq\pi/a$
and therefore is rotationally symmetric, and region II where $\pi/a\leq\left|\mathbf{k}\right|\leq\sqrt{3}\pi/a$
which consists of disconnected angular parts. Also as the three-momentum
integration is UV convergent, a small $a$ expansion of the integrand
can be performed. 
Using eq.~(\ref{eq:15}), the contribution from
region I to the $\mathbf{p}=0$ term in the Poisson sum is
\begin{eqnarray}
J_{LM}^{\left(I\right)}\left(\mathbf{p}=0\right)
& = & 
\frac{3\lambda}{\left(2\pi\right)^{4}}i^{L}
\int_{-\frac{\pi}{a}}^{\frac{\pi}{a}}dk_{4}\int_{0}^{\frac{\pi}{a}}dkk^{2}
\int d\Omega_{\hat{\mathbf{k}}}\frac{1}{\left(\hat{k}^{2}+m^{2}\right)^{2}}
\nonumber\\
&&\qquad\qquad  \times
\left[\ 
\int_{0}^{1}dy\ y^{2}\ j_{L}\left(aN\left|\mathbf{k}\right|y\right)\ \right]\ 
Y_{LM}\left(\Omega_{\mathbf{k}}\right)
\nonumber\\
& = & 
\frac{3\lambda}{16\pi^4}i^{L}
\left[J_{LM}^{LO}+J_{LM}^{NLO}+\mathcal{O}\left(1/N^{4}\right)\right]
\ \ ,
\label{eq:18}
\end{eqnarray}
where
\begin{eqnarray}
J_{LM}^{LO}
& = & 
2\sqrt{\pi}
\delta_{L,0}\delta_{M,0}
\int_{-\frac{\pi}{\Lambda a}}^{\frac{\pi}{\Lambda a}}\ 
dq_{4}\int_{0}^{\frac{\pi}{\Lambda
    a}}dqq^{2}\frac{1}{\left[q^{2}+q_{4}^{2}+m^{2}/\Lambda^{2}
\right]^{2}}
\ \int_{0}^{1}dy\ y^{2}\ j_0\left(qy\right)
\ \ \ ,
\nonumber
\end{eqnarray}
\begin{eqnarray}
J_{LM}^{NLO}
& = & 
\frac{1}{N^{2}}\int_{-\frac{\pi}{\Lambda a}}^{\frac{\pi}{\Lambda a}}
dq_{4}\int_{0}^{\frac{\pi}{\Lambda
    a}}dqq^{2}\frac{q^{4}}{\left[q^{2}+q_{4}^{2}+m^{2}/\Lambda^{2}\right]^{3}}
\nonumber\\
&& 
\qquad
\times
\left[
\frac{6\sqrt{\pi}}{5}\ \delta_{L,0}\ \delta_{M,0}\
\int_{0}^{1}dyy^{2}j_{0}\left(qy\right)
\right.\nonumber\\
&& \left.
\qquad
+\ \delta_{L,4}\left(\frac{2}{3}\sqrt{\frac{2\pi}{35}}\delta_{M,-4}
+\frac{4\sqrt{\pi}}{15}\delta_{M,0}
+\frac{2}{3}\sqrt{\frac{2\pi}{35}}\delta_{M,4}\right)\ \int_{0}^{1}dy\ y^{2}\ 
j_{4}\left(qy\right)\right]
\  ,
\label{eq:19}
\end{eqnarray}
with 
$q=\left|\mathbf{k}\right|/\Lambda$ and $q_{4}=k_{4}/\Lambda$.
The LO integral,  $J_{LM}^{LO}$, is convergent, while the NLO contribution, 
$J_{LM}^{NLO}$, while not convergent, is not divergent, but is of the 
form $\sin\left(N\pi\right)/N^2$.
This implies that they depend on the ratio of the two mass scales, $\Lambda$ and $m$, 
but without inverse powers of $a$. 
So as $a\rightarrow 0$, 
the LO $L=0$ operator makes an unsuppressed contribution to the $L=0$ matrix
element, while the contributions to this matrix element 
from the NLO rotational-symmetry violating $L=0$ and $L=4$ operators 
are suppressed by $1/N^{2}$.

A simple argument shows that contributions from integration region II, for
which  $\pi/a\leq\left|\mathbf{k}\right|\leq\sqrt{3}\pi/a$, 
are also
suppressed by $1/N^{2}$. 
After defining a new momentum variable 
$l_{\mu}=k_{\mu}a$ and $l^{2}=l_{1}^{2}+l_{2}^{2}+l_{3}^{2}$, the
$\mathbf{p}=0$ term of the Poisson sum in region II is
\begin{eqnarray}
J_{LM}^{\left(II\right)}\left(\mathbf{p}=0\right)
&  = & 
\frac{3\lambda}{16\pi^4}i^{L}
\int_{-\pi}^{\pi}dl_{4}\int_{\pi}^{\sqrt{3}\pi}dl\ l^{2}\ 
\int_{f\left(\Omega_{\mathbf{l}}\right)}d\Omega_{\mathbf{l}}
\nonumber\\
&&
\qquad\qquad
\frac{Y_{LM}\left(\Omega_{\mathbf{l}}\right)}{\left(4\sum_{\mu}\sin^{2}\left(l_{\mu}/2\right)
+a^{2}m^{2}\right)^{2}}
\int_{0}^{1}dy\ y^{2}\ j_{L}\left(Nly\right)
\ \ \ ,
\label{eq:20}
\end{eqnarray}
where $f\left(\Omega_{\mathbf{l}}\right)$ identifies the angular
region of integration, and whose parametric form does not matter for
this discussion. This region still exhibits cubic symmetry, and gives
rise to contribution to the $L=0,4,6,8,...$ operators. On the other
hand, the three-momentum integration is entirely located in the UV
as $a\rightarrow0$, and thus
\begin{equation}
\sin^{2}\left(l_{1}/2\right)+\sin^{2}\left(l_{2}/2\right)+\sin^{2}\left(l_{3}/2\right)
+\sin^{2}\left(l_{4}/2\right)\geq 1
\ \ \ \ .
\label{eq:21}
\end{equation}
Also, integration over the Bessel function brings in a factor of 
$\ -\cos\left(Nl\right)/(N^{2}l^{2})$, up to higher orders in $1/N$. 
So the integrand does not have any
singularities  in region II of the 
integration, and is bounded.  
As a result,
\begin{equation}
\left|J_{LM}^{\left(II\right)}\left(\mathbf{p}=0\right)\right|
\ \leq\ 
\frac{1}{N^{2}}
\frac{3\lambda}{(4\pi)^4}
\ \int_{-\pi}^{\pi}dl_{4}
\ \int_{\pi}^{\sqrt{3}\pi}dl
\ \int_{f\left(\Omega_{\mathbf{l}}\right)}d\Omega_{\mathbf{l}}
\ Y_{LM}\left(\Omega_{\mathbf{l}}\right)
\ \ \ ,
\label{eq:22}
\end{equation}
and consequently 
$J_{LM}^{\left(II\right)}\left(\mathbf{p}=0\right)$ 
itself is suppressed by $1/N^{2}$. 
This completes the discussion of the $\mathbf{p}=0$
term in the Poisson sum,
corresponding to a zero- momentum insertion of the continuum operator into the
loop diagram.
It then remains to determine the scaling of the  $\mathbf{p}\neq 0$ terms in the summation
in the large $N$ limit. 
The integral arising from the  $\mathbf{p}\neq 0$ terms
is, up to  numerical factors,
\begin{eqnarray}
{\cal I}_{\mathbf{p}\neq 0}
& \sim & 
\lambda\sum_{\mathbf{p}\neq 0}\int_{-\frac{\pi}{a}}^{\frac{\pi}{a}}\frac{d^{4}k}{\left(2\pi\right)^{4}}
\frac{1}{\left(\hat{k}^{2}+m^{2}\right)^{2}}\ 
Y_{L_{1}M_{1}}\left(\Omega_{\hat{k}}\right)\ 
Y_{L_{2}M_{2}}\left(\Omega_{\hat{p}}\right)
\nonumber\\ 
&& \qquad\qquad
\times\ 
\int_{0}^{1}dy\ y^{2}\ 
j_{L_{1}}\left(Na\left|\mathbf{k}\right|y\right)\ 
j_{L_{2}}\left(2\pi  N\left|\mathbf{p}\right|y\right)
\ \ \ \ .
\label{eq:23}
\end{eqnarray}
This integral is finite in UV, and  integrand can be expanded in powers of $a$,
giving  a leading contribution of 
\begin{eqnarray}
{\cal I}_{\mathbf{p}\neq 0}
& \sim & 
\lambda\sum_{\mathbf{p}\neq 0}\int_{-\frac{\pi}{\Lambda a}}^{\frac{\pi}{\Lambda
    a}}\frac{d^{3}q\ dq_{4}}{\left(2\pi\right)^{4}}\frac{1}{\left(q^{2}+q_{4}^{2}+m^{2}/\Lambda^{2}\right)^{2}}
\ Y_{L_{1}M_{1}}\left(\Omega_{\hat{\mathbf{q}}}\right)
\ Y_{L_{2}M_{2}}\left(\Omega_{\hat{p}}\right)
\nonumber\\ 
&& \qquad\qquad
\times\ 
\int_{0}^{1}dy\ y^{2}\ 
j_{L_{1}}\left(qy\right)
\ j_{L_{2}}\left(2\pi N\left|\mathbf{p}\right|y\right)
\ \ \ \ .
\label{eq:24}
\end{eqnarray}
A non-zero angular integration requires that $L_1=0$,
and the integral 
is suppressed at least by a factor of $1/N^{2}$ as
integration over the Bessel functions introduces a factor 
of $1/\left(2\pi N\left|\mathbf{p}\right|\right)^{2}$
up to a numerical coefficient and a bounded trigonometric function
at leading order in $1/N$. 
The next order term in the small $a$ expansion of the integrand
can be easily shown to bring in an additional  factor of $1/N^{2}$.
So one can see that the $\mathbf{p}\neq 0$ terms in the Poisson summation,
which give rise to non-continuum contributions to the two-point function
at one loop, are always suppressed by at least a factor of $1/N^{2}$.

The result of the one-loop calculation is promising: all the sub-leading
contributions that break rotational symmetry
are suppressed by $1/N^{2}$
compared to the leading $L=0$ continuum operator contribution to
the two-point function. 
A little investigation shows that this scaling
also holds to higher orders in $\lambda\phi^{4}$ theory. Suppose
that the operator is inserted into a  propagator inside an n-loop diagram
contributing to the two-point function. 
Considering the 
continuum part of the operator first, the leading term in the small $a$
expansion of the integrand gives rise to $2n$ propagators,
while the integration measure contributes $4n$ powers of momentum.
Although this appears to be logarithmically divergent, 
the spherical Bessel function contributes a factor of inverse three-momentum 
and either a sine or cosine of the three-momentum, rendering the diagram finite.
The same argument applies to the NLO term in the small $a$ expansion of
the integrand, resulting  in a $1/N^{2}$ suppression of the 
breaking of rotational invariance.
Insertion of the non-continuum operator in loop diagrams are also suppressed by
$1/N^2$ for similar reasons.

The interpretation of finite size scaling results presented in 
Refs.~\cite{CampostriniI,CampostriniII}
in terms of what has been observed in this section is now straightforward.
Near the critical point, the correlation length is the only relevant
physical scale in the problem, and tends to infinity. 
So as the critical point is approached, 
one does not probe the underlying lattice structure
as the correlation length becomes much larger than the lattice spacing, 
and extends over an
increasing number of point shells. 
In comparison,
inserting an operator which only probes distances of the order of
a physical scale that is much larger than the lattice spacing, 
resembles the physics near a rotational-invariant fixed point, and the same scaling
law for the non-rotational invariant operators is expected (in the
same theory) as the lattice spacing goes to zero.

\section{Operators in QCD
\label{sec:QCD}}
\noindent 
The necessity of introducing a gauge link to connect the fermionic
fields in a gauge invariant way, makes the discussion of the operator
and its renormalization more involved in gauge theories. 
The reason
is two-folded: firstly as is well known, perturbative
LQCD is ill-behaved as a result of non-vanishing tadpoles which diverge
in the  UV, making the small coupling series expansion of the operators
slowly convergent. The other difficulty is that as the operator is
smeared over many lattice sites, the links are necessarily extended links.
Thus, to analytically investigate the deviations
from a rotational invariant path, working with a well-defined
path on the grid is crucial. In this section, the strategies to deal
with these problems are discussed, and the scaling laws of different
operator contributions to the two-point function in 
QCD with an insertion of the smeared operator are deduced.

In position-space, perhaps the simplest gauge-invariant smeared operator of
quark bilinears is
\begin{equation}
\hat{\theta}_{L,M}\left(\mathbf{x};a,N\right)
\ =\ 
\frac{3}{4\pi N^{3}}\sum_{\mathbf{n}}^{\left|\mathbf{n}\right|\leq
  N}\overline{\psi}
\left(\mathbf{x}\right)U\left(\mathbf{x},\mathbf{x}+\mathbf{n}a\right)\psi\left(\mathbf{x}+\mathbf{n}a\right)
\ Y_{L,M}\left(\hat{\mathbf{n}}\right)
\ \ \ ,
\label{eq:25}
\end{equation}
where
\begin{equation}
U\left(\mathbf{x},\mathbf{x}+\mathbf{n}a\right)
\ =\ 
e^{ig\int_{\mathbf{x}}^{\mathbf{x}+\mathbf{n}a}\mathbf{A}\left(z\right)\cdot d\mathbf{z}}
\ =\ 
1+ig\int_{\mathbf{x}}^{\mathbf{x}+\mathbf{n}a}\mathbf{A}\left(z\right)\cdot d\mathbf{z}
+\mathcal{O}\left(g^{2}\right)
\ \ \ \ ,
\label{eq:26}
\end{equation}
where the actual path defining $U$ will be considered subsequently.
As the fermion operator is a spin singlet, $S=0$, 
the total angular momentum of this operator in the continuum is $J=L$.
One could also consider operators of the form
\begin{equation}
\hat{\theta}_{JL,M}^{\mu}\left(\mathbf{x};a,N\right)
\ =\ 
\frac{3}{4\pi N^{3}}
\sum_{\mathbf{n}}^{\left|\mathbf{n}\right|\leq   N}
\overline{\psi}
\left(\mathbf{x}\right)\ \gamma^\mu \ 
U\left(\mathbf{x},\mathbf{x}+\mathbf{n}a\right)
\psi\left(\mathbf{x}+\mathbf{n}a\right)
\ Y_{L,M}\left(\hat{\mathbf{n}}\right)
\ \ \ ,
\label{eq:25b}
\end{equation}
which can be used to form operators with $J=L+1, L, L-1$.  
It is clear that
the set of operators with angular momentum $J$ will mix under renormalization,
but the vector nature of QCD precludes mixing between the
$\overline{\psi}\psi$ and $\overline{\psi}\gamma^\mu \psi$ operators in the
chiral limit.
However to capture the main features
of operator mixing in the continuum limit of LQCD, it suffices to
work with the simplest operator, in eq.~(\ref{eq:25}).
\begin{figure}
\begin{centering}
\includegraphics[scale=0.12]{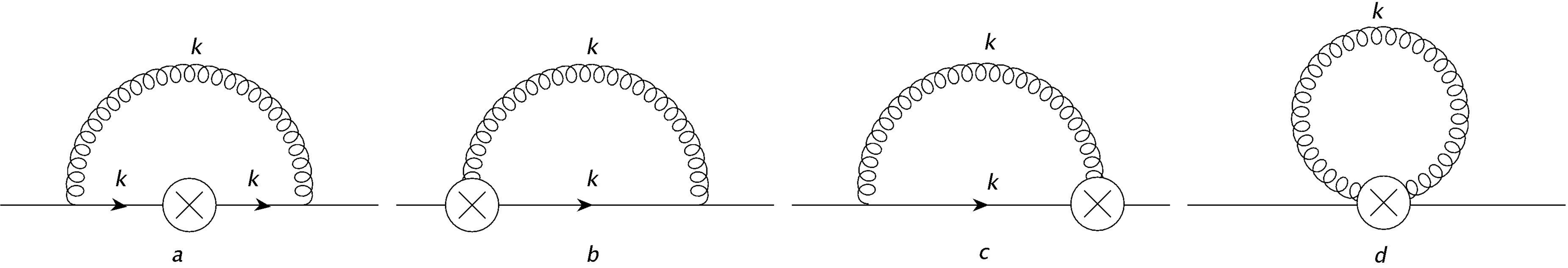}
\par\end{centering}
\caption{{\small 
One-loop QCD corrections to the fermionic two-point function
with an insertion of 
$\hat{\theta}_{L,M}$, given in eq.~(\ref{eq:25}),
at zero external momentum}}
\label{fig:qcdOneLoop}
\end{figure}
At tree-level, the contributions of this operator away from the continuum limit
scale in the same way as in the scalar theory, 
with contributions that violate rotational invariance 
suppressed by $\sim 1/N^2$.

Let us first discuss the one-loop renormalization of the operator
in the continuum. There are three one-loop diagrams contributing to
the operator renormalization as shown in fig.~\ref{fig:qcdOneLoop}. 
The diagram in
fig.~\ref{fig:qcdOneLoop}a 
results from inserting the leading order term in the small
coupling expansion of the operator in the loop. At zero external momentum
this diagram is
\begin{eqnarray}
\Gamma^{(5a)} & \sim & 
-T^{a}T^{a}\frac{3ig^{2}}{4\pi}\int_{0}^{1}dyy^{2}
\int d\Omega_{\mathbf{y}}\int\frac{d^{4}k}{\left(2\pi\right)^{4}}
\frac{\gamma_\alpha
\left(ik_{\mu}\gamma^{\mu}+m\right)^{2}
\gamma^\alpha
}{\left(k^{2}+m^{2}\right)^{2}k^{2}}
e^{iNa\mathbf{k}\cdot\mathbf{y}}Y_{LM}\left(\Omega_{\mathbf{y}}\right)
\ \ \ ,
\label{eq:27}
\end{eqnarray}
which is clearly convergent in the UV. 
Also it contains $L=0$
as well as $L=1$ operator as can be seen from the angular
part of the integral
\begin{eqnarray}
&&
\sum_{L^{\prime},M^{\prime}}\int d\Omega_{\mathbf{y}}d\Omega_{\mathbf{k}}
\left[f_{1}\left(k^{2},m,k_{4}\right)+f_{2}\left(k^{2},m,k_{4}\right)\mathbf{k}\cdot\vec{\mathbf{\gamma}}\right]
\ Y_{L^{\prime}M^{\prime}}\left(\Omega_{\mathbf{k}}\right)
\ Y^{*}_{L^{\prime}M^{\prime}}\left(\Omega_{\mathbf{y}}\right)
\ Y_{LM}\left(\Omega_{\mathbf{y}}\right)
\nonumber\\
&&
\ =\ 
\sqrt{4\pi}f_{1}\left(k^{2},k_{4},m\right)\delta_{L,0}\delta_{M,0}
\nonumber\\
&&
\qquad 
\ +\ 
\sqrt{\frac{4\pi}{3}}f_{2}\left(k^{2},k_{4},m\right)\left|\mathbf{k}\right|
\delta_{L,1}\left[\gamma_{1}\left(\frac{\delta_{M,-1}-\delta_{M,1}}{\sqrt{2}}\right)
+i\gamma_{2}\left(\frac{\delta_{M,-1}+\delta_{M,1}}{\sqrt{2}}\right)+\gamma_{3}\delta_{M,0}
\right]
\ \ \ ,
\nonumber\\
&&
\label{eq:28}
\end{eqnarray}
where $f_{1}$ and $f_{2}$ are some functions of their arguments.
One can check however that as $m/\Lambda\rightarrow0$ (the chiral
limit), the contribution to the $L=1$ operator is suppressed by the
quark mass.

The diagrams in fig.~\ref{fig:qcdOneLoop}b comes from the next term in the expansion
of eq. (\ref{eq:26}). It is straightforward to show that the Feynman
rule for the one-gluon vertex with zero momentum insertion into the
operator is
\begin{eqnarray}
V_g^\lambda & = & 
\frac{3}{4\pi N^{3}}\sum_{\mathbf{n}}^{\left|\mathbf{n}\right|\leq  N}
 g a n^\lambda
\ \frac{1}{\left(\mathbf{p}-\mathbf{p^{\prime}}\right)\cdot\mathbf{n}a}
\left(e^{i\left(\mathbf{k}+\mathbf{p^{\prime}}\right)\cdot \mathbf{n}a}
-e^{i\mathbf{p^{\prime}\cdot}\mathbf{n}a}\right)\delta^{4}\left(p-p^{\prime}-k\right)
\ Y_{L,M}\left(\hat{\mathbf{n}}\right)
\ \ \ \ ,
\qquad
\label{eq:29}
\end{eqnarray}
where the radial path between points $\mathbf{x}$ and $\mathbf{x}+\mathbf{n}a$
is taken in evaluating the link integral, $p$ and $p^{\prime}$ are
the momenta of incoming and outgoing fermions respectively, 
$\lambda$ is the Lorentz-index of the gluon field,
and $k$
is the momentum of the gluon coming out of the vertex. Note that in
principle, any path between points $\mathbf{x}$ and $\mathbf{x}+\mathbf{n}a$
can be taken in the above calculation, but if one is interested in
deviations of the renormalized lattice operator from the rotational
invariance compared to the continuum operator, a path between two
points should be chosen in the continuum in such a way that it respects
rotational invariance explicitly. 
Any path other than the radial path, on the other hand, is equivalent to
infinitely many other paths resulting
from rotated versions of the original path around the radial path.
To reveal rotational invariance at the level of the
continuum operator, an averaging over these infinite copies of the
path is needed, and this makes the calculation of the link more involved.

Now at zero external momentum, using expression (\ref{eq:29}) with
$p=0$, the contribution from the second and third diagrams in
fig.~\ref{fig:qcdOneLoop}b is
\begin{eqnarray}
\Gamma^{(5b,5c)}
& \sim & 
- T^{a}T^{a}
\frac{3 g^2}{2\pi}\int_{0}^{1}dyy^{2}\int d\Omega_{\mathbf{y}}
\int\frac{d^{4}k}{\left(2\pi\right)^{4}}
\frac{
i {\bf k}\cdot {\bf y} 
+m \mathbf{y}\cdot\vec{\gamma}
}{\left(k^{2}+m^{2}\right)k^{2}}
\nonumber\\
&&
\qquad
\qquad
\times\frac{1}{\mathbf{k}.\mathbf{y}}\left(e^{iNa\mathbf{k}\cdot\mathbf{y}}-1\right)
\ Y_{LM}\left(\Omega_{\mathbf{y}}\right)
\ \ \ \ .
\label{eq:30}
\end{eqnarray}
As is evident, because of a non-oscillatory contribution to the operator,
there is a logarithmically divergent piece from the above integration
contributing to the $L=0$ operator, which along with the logarithmic
divergent  contribution  
from wavefunction renormalization, contributes
to the anomalous dimension of the operator. Also the angular integration
of the above expression: 
\begin{eqnarray}
&& \int d\Omega_{\mathbf{y}}d\Omega_{\mathbf{k}}
\left[1+\frac{\mathbf{y}\cdot\vec{\gamma}}{i\mathbf{k}\cdot\mathbf{y}} m \right]
\left(e^{iNa\mathbf{k}\cdot\mathbf{y}}-1\right)
\ Y_{LM}\left(\Omega_{\mathbf{y}}\right)
\nonumber\\
&&\ =\ 
\int d\Omega_{\mathbf{y}}
\left[g_{1}\left(Nay\left|\mathbf{k}\right|\right)
+g_{2}\left(Nay\left|\mathbf{k}\right|\right) m \mathbf{y}\cdot\vec{\gamma}\right]
\ Y_{LM}\left(\Omega_{\mathbf{y}}\right)
\ \ \ ,
\label{eq:31}
\end{eqnarray}
indicates that as before, in addition to $L=0$ operator, an $L=1$
contribution is present which is finite at UV, and can be shown to
vanish for $m/\Lambda\rightarrow0$. $g_{1}$ and $g_{2}$ are some
functions of their arguments whose explicit forms do not matter for
this discussion.

\begin{figure}
\begin{centering}
\includegraphics[scale=0.17]{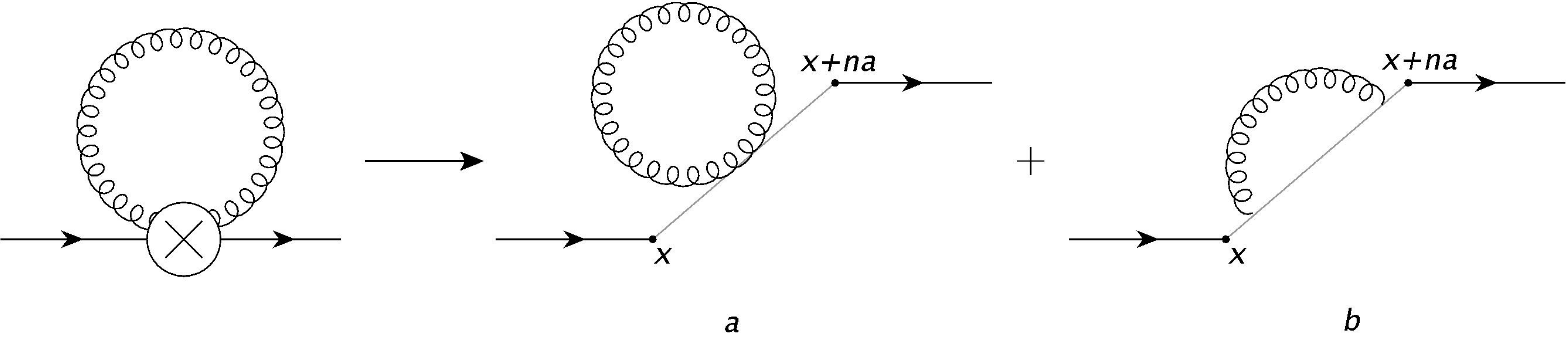}
\par\end{centering}
\caption{{\small The tadpole contribution consists of the conventional tadpole
diagram (a), which vanishes when using a mass-independent regulator in the
continuum (such as dimensional  regularization), 
as well as the diagram
shown in (b) which is of the order of $\alpha_{s}/\left|\mathbf{\Delta x}\right|^{2}$,
where $\mathbf{\Delta x}$ is the distance between two gluon vertices.}}
\label{fig:tadpoles}
\end{figure}
The last diagram in fig.~\ref{fig:qcdOneLoop} corresponds to the $\mathcal{O}\left(g^{2}\right)$
term in the small coupling expansion of the gauge link. 
It contains
the tadpole of the continuum theory whose value depends in general
on the regularization scheme.
For example, by using a hard momentum cutoff which is matched easily
with the lattice regularization, it diverges quadratically. However,
it is not hard to see that in dimensional regularization which respects
the full rotational symmetry of the continuum, it vanishes in $d=4$,
therefore it does not contribute to the renormalization of the continuum
operator. 
But the fourth diagram in fig.~\ref{fig:qcdOneLoop} does not only include the
conventional tadpoles, fig.~\ref{fig:tadpoles}a, it also contains the diagram where
a gluon is emitted by the Wilson line inside the operator and then
absorbed at another point on the Wilson line, fig.~\ref{fig:tadpoles}b as a consequence
of the matter fields being separated by a distance $\mathbf{n}a$.
It is straightforward to show this diagram is convergent, and scales
by $\alpha_{s}/\left|\mathbf{\Delta x}\right|^{2}$ where $\mathbf{\Delta x}$
is the distance between two gluon vertices and $\alpha_{s}$ is evaluated
at the energy scale of the order of $1/\left|\mathbf{\Delta x}\right|$.
This completes the qualitative discussion of the operator renormalization
and mixing at one-loop order in the continuum.

Let us start the discussion of the lattice operator by assuming that its
definition is still given by eq.~(\ref{eq:25}). However, this can
be shown to be a naive definition of the operator on the lattice.
The reason is implicit in the discussion of tadpoles given above.
Although  tadpoles are absent from the operator renormalization
in the continuum,
on the lattice, they are non-vanishing, and result in large
renormalizations, as can be seen  in perturbative lattice QCD calculations. 
As was suggested
long ago by Lepage and Mackenzie \cite{Lepage}, to make the perturbative
expansion of the lattice quantities well-behaved, and to define an
appropriate connection between the lattice operators and their continuum
counterparts, one can remove tadpoles from the expansion of the lattice
operators in a non-perturbative manner by dividing the gauge link
by its expectation value in a smooth gauge,
\begin{equation}
U\left(x,x+a\hat{\mu}\right)
\rightarrow
\frac{1}{u_{0}}U\left(x,x+a\hat{\mu}\right)
\ \ \ \ ,
\label{eq:32}
\end{equation}
where a simpler, gauge invariant choice of $u_{0}$ uses the measured
value of the plaquette in the simulation,
 $u_{0}\equiv\left\langle \frac{1}{3}{\rm Tr}\left(U_{plaq}\right)\right\rangle
 ^{1/4}$.
There remains still another issue regarding the tadpole contributions
to the smeared operator which is not fully taken care of by the simple
single-link improvement procedure explained above. The operator introduced in
eq.~(\ref{eq:25}) is smeared over several lattice sites, and as a
result includes extended links. As will be explained shortly, in
spite of $\mathcal{O}\left(\alpha_{s}\right)$ corrections due to
tadpoles from a single link, 
there is an $\mathcal{O}\left(N\alpha_{s}\right)$
enhancement due to the tadpoles from the extended link with length $\sim Na$. 
So although a non-perturbative tadpole improvement
could introduce non-negligible  statistical errors, this improvement is crucial,
otherwise the relation between the lattice smeared operator and
the corresponding continuum operator is somewhat obscure.
\begin{figure}[t]
\begin{centering}
\includegraphics[scale=0.15]{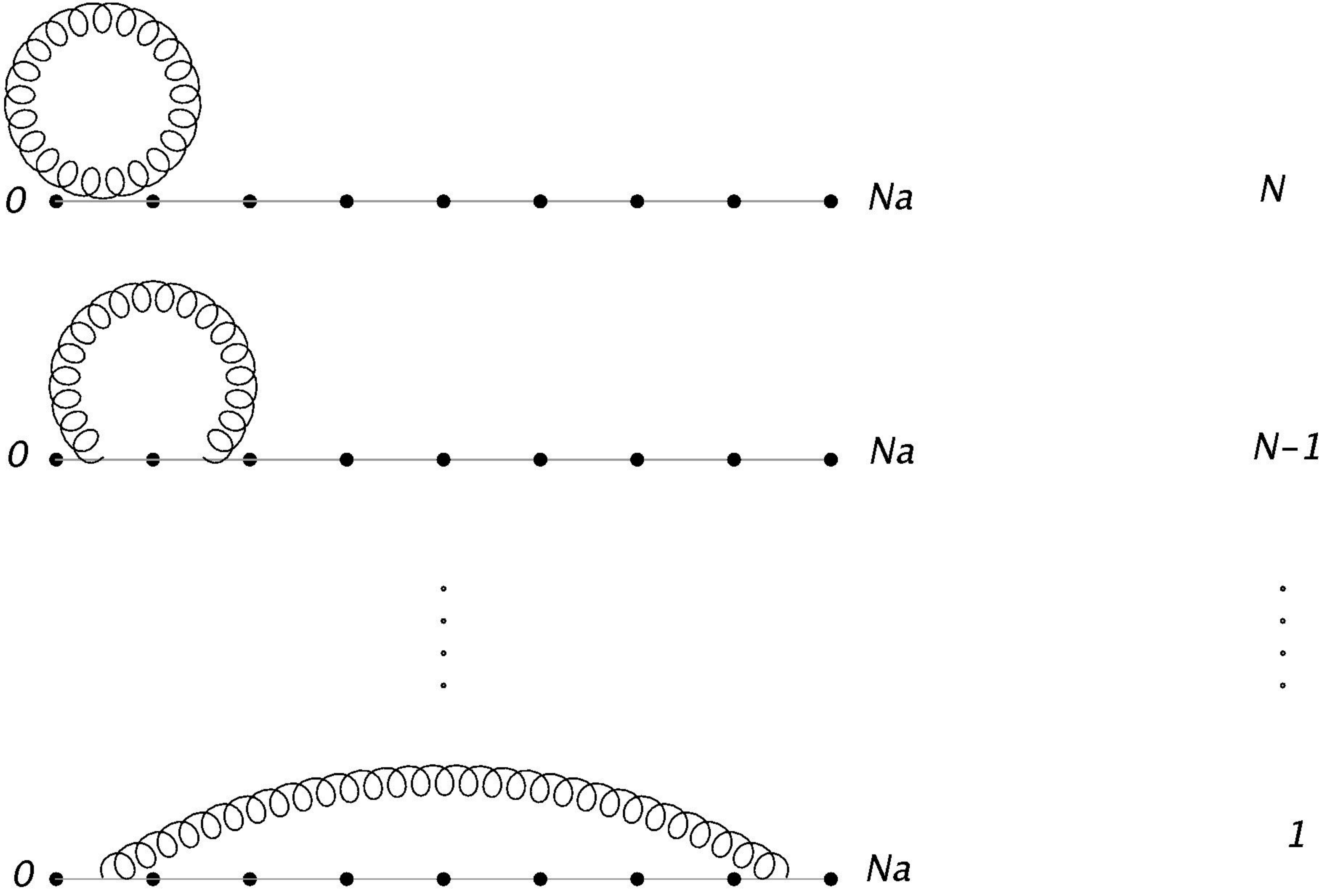}
\par\end{centering}
\caption{{\small Tadpole diagrams contributing to the smeared operator at one-loop
order. Shown in the right are the number of diagrams of each type.}}
\label{fig:tadpoleextended}
\end{figure}

The reason for the $\mathcal{O}\left(N\alpha_{s}\right)$ enhancement 
of tadpoles from the extended links
can be illustrated by working out a particular example. Suppose that
the link is extended between points $\mathbf{x}$ and $\mathbf{x}+Na\hat{e}_{1}$
entirely along the $1$ axis. 
Then in order to make a tadpole, not
only can each gauge field be contracted with the other gauge field belonging to
the same elementary link, but also it can be contracted with a gauge field
from one of the remaining $N-1$ elementary links 
(see fig.~\ref{fig:tadpoleextended}). 
Note that each diagram in fig.~\ref{fig:tadpoleextended} 
comes with a multiplicity of $N-m$, where $m$ is the number of
links between the contracted gluonic vertices.
At LO  in $a$, the corresponding contribution from the extended tadpole (ET) is of the form
\begin{equation}
\Gamma^{(ET)}
\sim
 \alpha_{s}a^{2}\int_{-\frac{\pi}{a}}^{\frac{\pi}{a}}d^{4}k
\ \frac{e^{imak_{1}}}{k_{1}^{2}+k_{2}^{2}+k_{3}^{2}+k_{4}^{2}}
\ \sim\ 
\frac{\alpha_{s}}{m^{2}}
\ \ \ \ ,
\label{eq:33}
\end{equation}
from which
the contribution from all the diagrams in fig.~\ref{fig:tadpoleextended} can be obtained,
\begin{equation}
\sum_{m=1}^{N-1}(N-m)\frac{\alpha_{s}}{m^{2}}
\ =\ 
\mathcal{O}\left(N\alpha_{s}\right)
\ \ \ .
\label{eq:34}
\end{equation}
Note that the $m=0$ term, corresponding to the first diagram in
fig~\ref{fig:tadpoleextended}, 
has been excluded from the above sum 
as it is just the single link tadpole contribution.
Given that there are N single links, the total contribution from single link
tadpoles is  $\mathcal{O}\left(N\alpha_{s}\right)$ as well.

Another issue with the extended links is the fact that without
tadpole improvement, breakdown of rotational symmetry occurs
at $\mathcal{O}\left(N\alpha_{s}\right)$. 
The reason is that without tadpole improvement of the extended links, 
contributions from the different 
$A_{1}$ irreps in a given point shell are normalized differently.
For example, there are more tadpole diagrams at 
$\mathcal{O}\left(g^{2}\right)$ contributing to an extended 
link between points $\left(0,0,0\right)$ and $\left(2,2,1\right)$ (six single links)
than to an extended link between points $\left(0,0,0\right)$ and
$\left(3,0,0\right)$
(three single links)
although both points belong to the same point shell  
(i.e. have the same separation in position space). 
This fact magnifies the necessity of tadpole improvement as well as providing  
a prescription for an appropriate improvement of an extended link. 
As the
expectation value of a link belonging to a given $A_{1}$ irrep in
a given shell is in general different from the expectation value of
the link belonging to another $A_{1}$ irrep in the same shell, one
needs to redefine the link in a given irrep by dividing it by its
expectation value in the same irrep,
\begin{equation}
U_{A_{1}^{i}}\left(x,x+a\mathbf{n}\right)
\ \rightarrow
\ \frac{1}{u_{A_{1}^{i}}}U_{A_{1}^{i}}\left(x,x+a\mathbf{n}\right)
\ \ \ \ ,
\label{eq:35}
\end{equation}
where $u_{A_{1}^{i}}=\left\langle U_{A_{1}^{i}}\left(x,x+a\mathbf{n}\right)\right\rangle $,
and the $A_{1}^{i}$'s are different $A_{1}$ irreps belonging to the
$n^{2}$-shell. 
With this prescription for tadpole improvement of the extended links, the
renormalized operator is 
assured
to be safe from large rotational invariance breaking effects of the
order of $\mathcal{O}\left(N\alpha_{s}\right)$. 
With this new definition
of the gauge link, eq. (\ref{eq:25}) is now a well-defined lattice
operator with an appropriate continuum limit which can be used
in our subsequent analysis.

As the cancellation of the tadpole diagram is assured by the new definition
of the operator, there are only  three one-loop diagrams  that contribute
to the renormalization of the lattice operator. 
The first
diagram in fig.~\ref{fig:qcdOneLoop} corresponds to the following loop integral at zero
external momentum for Wilson fermions,
\begin{eqnarray}
\Gamma^{(5a)}
& \sim & 
\left(ig\right)^{2}T^{a}T^{a}\frac{3}{4\pi N^{3}}\sum_{\mathbf{n}}
 \int_{-\frac{\pi}{a}}^{\frac{\pi}{a}}\frac{d^{4}k}{\left(2\pi\right)^{4}}
 e^{i\mathbf{k}\cdot\mathbf{n}a}
 \left[\gamma_{\rho}\cos\left(\frac{k_{\rho}a}{2}\right)-ir\sin\left(\frac{k_{\rho}a}{2}\right)\right]
\nonumber\\
&&
\times
\ \left(\frac{-i\sum_{\mu}\gamma_{\mu}\frac{\sin\left(k_{\mu}a\right)}{a}+M\left(k\right)}
{\sum_{\mu}\frac{\sin^{2}\left(k_{\mu}a\right)}{a^{2}}+M\left(k\right)^{2}}\right)^{2}
\left[\gamma^{\rho}\cos\left(\frac{k^{\rho}a}{2}\right)-ir\sin\left(\frac{k^{\rho}a}{2}\right)\right]
\nonumber\\
&&
\times\ 
\frac{i}{\frac{4}{a^{2}}\sum_{\nu}\sin^{2}\left(\frac{k_{\nu}a}{2}\right)}
\ Y_{LM}\left(\Omega_{\mathbf{n}}\right)
 \ \ \ , 
\label{eq:36}
\end{eqnarray}
where $M\left(k\right)\equiv M+2r/a\sum\limits_{\mu}\sin^{2}\left(k_{\mu}a/2\right)$,
and $r$ is the Wilson parameter. Clearly at LO in
the lattice spacing, one recovers the corresponding diagram with the
insertion of the continuum operator, eq.~(\ref{eq:27}), and 
so it contributes to both the $L=0$ and $L=1$ operators. 
Note that although the integration region
is not rotationally symmetric like the continuum integral, the convergence
of integral at UV ensures that the contributions from non-rotationally
symmetric integration region II, defined in section~\ref{sec:Scalar},
are suppressed by additional  powers of $1/N$ compared to the rotational
invariant region I:
\begin{eqnarray}
\delta\Gamma^{(5a)}
& \sim & 
-ig^{2}T^{a}T^{a}\frac{3
    i^{L}}{16\pi^{4}}
\int_{-\pi}^{\pi}dl_{4}\int_{\pi}^{\sqrt{3}\pi}dl\ l^{2}\ 
\int_{f\left(\Omega_{\mathbf{l}}\right)}d\Omega_{\mathbf{l}}\ 
\frac{\left(il_{\mu}\gamma^{\mu}+ma\right)^{2}}{\left(l^{2}+m^{2}a^{2}\right)^{2}l^{2}}
\nonumber\\
&&
\qquad
\qquad
\times\ Y_{LM}\left(\Omega_{\mathbf{l}}\right)\left[\int_{0}^{1}dyy^{2}j_{L}\left(Nly\right)\right]
\ \ \ ,
\label{eq:37}
\end{eqnarray}
where: $l_{\mu}=k_{\mu}a$ and $l^{2}=l_{1}^{2}+l_{2}^{2}+l_{3}^{2}$.
The integrand is clearly convergent, and the integration region is
entirely in the UV, and  so the only dependence on $a=1/(\Lambda N)$ comes from
the integration over the Bessel function, giving a LO contribution 
proportional to $1/N^{2}$. 
However,
the first sub-leading contribution from this
diagram scales as  $\sim \alpha_{s}/N$ for Wilson fermions instead
of $\sim \alpha_{s}/N^{2}$. 
The reason is that the small $a$
expansion of the integrand in eq.~(\ref{eq:36}) includes terms at
$\mathcal{O}\left(a\right)$ which is proportional to the Wilson
parameter. 
The integrand
scales as $\sim 1/k^{3}$ multiplied by the spherical Bessel function
in the UV which still gives rise
to a convergent four-momentum integration for any value of $L$,
\begin{equation}
\delta\Gamma^{(5a,r)}
\sim a\int d^{4}k\frac{1}{k^{3}}\left[\int_{0}^{1}dy\ y^{2}
\ j_{L}\left(Naky\right)\right]\sim a\Lambda=\frac{1}{N}
\ \ \ .
\label{eq:38}
\end{equation}
These contributions are rotational invariant, and will be included in the
renormalization 
$Z$-factor of the operator when matching the lattice operator with its
continuum counterpart. 
Further,
the integrals that appear at $\mathcal{O}\left(a^{2}\right)$
in an expansion of eq.~(\ref{eq:36}) are also convergent,
and the terms containing rotational invariance breaking
contributions are suppressed by $1/N^{2}$. This completes discussion
of the first one-loop diagram of fig.~\ref{fig:qcdOneLoop}.

The second diagram contains the one-gluon vertex operator, and requires
evaluating a line  integral over the path on the grid defining the extended
link. 
As was pointed
out in the discussion of the path in the continuum, in general any
path can be chosen in evaluating the operator both in the continuum
or on the lattice, but requiring the recovery of rotational symmetry
at the level of the operator means that the extended link has to exhibit
rotational symmetry in the continuum limit. 
As already discussed, the simplest rotational
invariant path in the continuum is the radial path between the points,
so it makes sense to try to construct a path on the grid which remains
as close as possible to the radial path between points $x$ and $x+\mathbf{n}a$
as it passes through the lattice sites. One might expect though that
choosing a path in continuum which is the same as its lattice counterpart
is a more legitimate choice. One example of such a path is an $L$-shaped
path. However, it is not hard to verify that the $L$-shaped link does not
restore rotational invariance in the continuum limit as 
the continuum path explicitly breaks rotational symmetry.
So the problem of evaluating the one-gluon vertex of the smeared operator
is reduced to finding the closest path to the straight line on the
grid. 
In a lattice calculation, one can, in principle, construct an algorithm
which finds a path on the three-dimensional grid in such a way that
the area between the path and the rotational invariant radial path
is a minimum. 
One such algorithm has already been used in Ref.~\cite{Meyer}
to construct a path that follows the straight line between sites A and B as
closely as possible, by forming a diagonal link at each step which has
the maximum projection onto the vector $\overrightarrow{AB}$. 
By
this construction of ``super-links'', the authors have been able
to form arbitrary (approximate) rotations of the Wilson loops, therefore
constructing glueball operators which project onto a definite spin $J$
in the continuum limit. 
However, 
the analytic form of the super-link has not been given.
In appendix \ref{app:linksongrid}, a method to evaluate the link on such a path
is illustrated with a small number of examples. 
For the following discussion
however, a particular example has been considered which encapsulates
the essential features of the recovery of the rotational path, and
gives us an idea how to deal with the general case.

\begin{figure}
\begin{centering}
\includegraphics[scale=0.17]{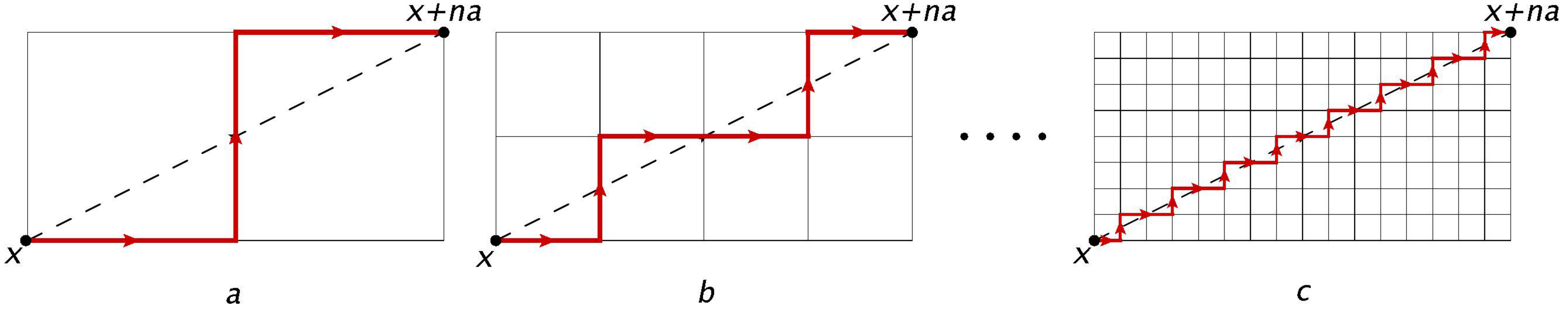}
\par\end{centering}
\caption{{\small a) The link between points $x$ and $x+\mathbf{n}a$ for $\mathbf{n}=\left(2,1,0\right)$
which remains as close as possible to the diagonal link, b) The link
between the same points for $\mathbf{n}=2\left(2,1,0\right)$ which
consists of two separate links of part a) with the lattice spacing
being halved, c) The link for $\mathbf{n}=2^{K}\left(2,1,0\right)$
which consists of $2^{K}$ separate links of part a) with the  lattice
spacing divided by $2^{K}$.}}
\label{fig:links}
\end{figure}
Suppose that the link connects points $x$ and $x+\mathbf{n}a$ on
a cubic lattice where
$\mathbf{n}=\frac{a_{0}}{a}\left(Q,1,0\right)$,
and $a_{0}=2^{K}a$. As usual $a$ denotes the lattice spacing, and
$Q$ is an arbitrary integer. The continuum limit is recovered when
the integer $K$ tends  to infinity for a finite value of $a_{0}$.
Then as is shown in appendix \ref{app:linksongrid}, for a path which is symmetric under
reflection about its midpoint and remains as close as possible to
the vector $\mathbf{n}a$ (see fig.~\ref{fig:links}), the $\mathcal{O}\left(g\right)$
term in the momentum-space expansion of the link has the following
form
\begin{eqnarray}
U^{\left(1g\right)}\left(q\right)
 & = &  
ig\frac{a_{0}}{2^{K}}e^{i\mathbf{q}\cdot\mathbf{n}a/2}
\frac{\sin\left(\frac{\mathbf{q}\cdot\mathbf{n}a}{2}\right)}{
\sin\left(\frac{\mathbf{q}\cdot\mathbf{n}a}{2^{K+1}}\right)}
\left[A_{y}\left(q\right)
\right.
\nonumber\\
&&
\left.
\qquad\qquad\qquad
+2A_{x}\left(q\right)\frac{\sin\left(Qq_{x}a_{0}/2^{K+2}\right)}{
\sin\left(q_{x}a_{0}/2^{K+1}\right)}\cos\left(\frac{Qq_{x}a_{0}}{2^{K+2}}+\frac{q_{y}a_{0}}{2^{K+1}}\right)
\right]
\ \ .
\label{eq:39}
\end{eqnarray}
As $K\rightarrow\infty$ limit which corresponds to $a\rightarrow 0$, one
obtains
\begin{eqnarray}
U^{\left(1g\right)}\left(q\right)
 & = &  
2ige^{i\mathbf{q}\cdot\mathbf{n}a/2}\frac{\sin\left(
\frac{\mathbf{q}\cdot\mathbf{n}a}{2}\right)}{\mathbf{q}\cdot\mathbf{n}a}
\left[
\mathbf{A}\cdot\mathbf{n}a+\frac{a^{2}}{24}
\left(q_{x}Q+q_{y}\right)^{2}\mathbf{A}\cdot\mathbf{n}a
\right.
\nonumber\\
&& 
\left.
\qquad\qquad\qquad
-\frac{a^{2}}{24}QA_{x}a_{0}\left(q_{x}^{2}\left(Q^{2}-1\right)
+3Qq_{x}q_{y}+3q_{y}^{2}\right)
+\mathcal{O}\left(a^{4}\right)
\right]
 \ \ \ ,
\label{eq:40}
\end{eqnarray}
recovering the  continuum link, given in  eq.~(\ref{eq:29}),
and contains broken rotational invariance contributions which are
suppressed by $\sim {\cal O}(a^{2})$. This
scaling has been shown in appendix \ref{app:linksongrid} to hold for vectors $\mathbf{n}$
of the forms: $\frac{a_{0}}{a}\left(Q,1,1\right)$, $\frac{a_{0}}{a}\left(Q,Q,1\right)$
and $\frac{a_{0}}{a}\left(Q,Q,Q\right)$ as well.

Let us now examine
how the insertion of this contribution from the operator  modifies the scaling
of the rotational invariance violating operators at one-loop.
The contribution from the second diagram in fig.~\ref{fig:qcdOneLoop} with
the insertion of this vertex can be calculated order by order in small
$a$ by expanding the vertices and propagators as before. 
At the LO one gets
\begin{eqnarray}
\Gamma^{(5b)}
& \sim & 
-ig^{2}T^{a}T^{a}\frac{3}{4\pi
  N^{3}}\sum_{\mathbf{n}}\int_{-\frac{\pi}{a}}^{\frac{\pi}{a}}
\frac{d^{4}k}{\left(2\pi\right)^{4}}\frac{ik_{\mu}\gamma^{\mu}+m}{\left(k^{2}+m^{2}\right)k^{2}}
\frac{e^{i\mathbf{k}\cdot\mathbf{n}a}-1}{i\mathbf{k}\cdot\mathbf{n}a}
\ Y_{LM}\left(\Omega_{\mathbf{n}}\right)
\nonumber\\
&&
\times\left[a\mathbf{n}\cdot\vec{\gamma}
+\frac{a^{2}}{24}\left(k_{x}Q+k_{y}\right)^{2}a\mathbf{n}\cdot\vec{\gamma}
-\frac{a^{2}}{24}Q\left(k_{x}^{2}\left(Q^{2}-1\right)+3Qk_{x}k_{y}+3k_{y}^{2}\right)\gamma_{x}a_{0}\right]
\ \ \ \ .
\qquad
\label{eq:41}
\end{eqnarray}
Clearly, after adding the contribution from the third diagram in fig.~\ref{fig:qcdOneLoop}, the LO contribution from the above expression,
the first term in the bracket of eq.~(\ref{eq:41}),
 recovers
the results obtained previously for the insertion of the continuum
operator, up to suppressed contributions from the integration region
II, as discussed before. Therefore this term contributes to the $L=0$
operator with a logarithmically divergent coefficient, which along
with the wavefunction renormalization contributes to the anomalous
dimension of the lattice operator. Note that the wavefunction renormalization
gives rise to a logarithmically divergent contribution to the $L=0$
operator at LO  in the lattice spacing, recovering the continuum result,
and the sub-leading contributions are suppressed at least by $a=1/(N\Lambda)$
for Wilson fermions. This term also contains and $L=1$ operator which
is proportional to $m$, and vanishes in the chiral limit.

The second  term in the bracket of  eq.~(\ref{eq:41}) 
is ${\cal O}(a^2)$, and can be written as
\begin{eqnarray}
\delta\Gamma^{(5b,5c),2}
& =  & \
-i{g^2 a^2\over 8\pi N^3}\ 
T^{a}T^{a}
\
\sum_{\mathbf{n}}
\int_{-\frac{\pi}{a}}^{\frac{\pi}{a}}\frac{d^{4}k}{\left(2\pi\right)^{4}}
\left[1+\frac{m}{i\mathbf{k}\cdot\mathbf{n}a}a\mathbf{n}\cdot\vec{\gamma}\right]
\frac{e^{i\mathbf{k}\cdot\mathbf{n}a}-1}{\left(k^{2}+m^{2}\right)k^{2}}
\nonumber\\
&&
\qquad
\qquad
\qquad
\qquad
\qquad
\qquad
\times\left(k_{x}Q+k_{y}\right)^{2}Y_{LM}\left(\Omega_{\mathbf{n}}\right)
\nonumber\\
& \sim & \mathcal{O}\left(g^{2}a^{0}\right)
\ \ \ \ .
\label{eq:42}
\end{eqnarray}
This scaling arises as a result of the UV divergence of the non-oscillatory
contribution to the integral and is entirely a UV effect.  For this term there
is no dependence upon ${\bf n}$ and as such the factor of $N^{-3}$ is canceled
by a corresponding $N^3$ from the sum.
Terms proportional
to the mass are convergent in the UV, and as such are suppressed by $a^{2}$
in the continuum limit.

The last term in the above expression eq. (\ref{eq:41}) contains
rotational breaking contributions. 
It is multiplied by an explicit factor of 
$a^{2}$, 
but as seen in the previous term, 
the power divergence of the non-oscillatory part of the integral gives rise to 
an overall scaling of $\mathcal{O}\left(g^{2}\right)$. 
This completes the discussion
of the one-loop corrections to the lattice operator for the specific displacement
vector $\mathbf{n}a$ used above. 
It is also straightforward to check
the obtained scaling of different terms for other choices of the vector
$\mathbf{n}a$. 
In general, sub-leading contributions to the continuum link
are ${\cal O}(a^{2})$, and
so by dimensional analysis it 
has an associated factor of momentum squared. 
On the other hand, it always contains a non-oscillatory
term, and  as a result, the non-continuum
contributions  and the violations of rotational symmetry 
scale as $\mathcal{O}\left(\alpha_{s}\right)$.

Given the discussion of the previous paragraphs, we naively conclude that the
rotational symmetry breaking scales as $\sim {\cal O}(\alpha_s)$ in the continuum limit.
It is the one-gluon vertex associated with the smeared-operator that is
dominating this behavior, with the contributions from other diagrams
scaling as $\sim \alpha_s/N$ for Wilson fermions (eq.~(\ref{eq:36}) and eq.~(\ref{eq:37}))
and $\alpha_s/N^2$ from the other loop
diagrams compared with $\sim 1/N^2$ from the tree-level matching.
However, this scaling can be further improved by smearing the gauge-field.
The ${\cal O}(\alpha_s)$ contributions are due to the  explicit factor of $a^2$ being
compensated by a quadratic loop divergence, $\left(\pi/a\right)^2$, rendering a
suppression by only the coupling in the continuum limit, analogous to the
impact of tadpole diagrams. 
However, by smearing the gluon field over a volume of radius 
$1/\Lambda_g = a N_g$~\footnote{We have
distinguished the smearing radius of the operator, $N$, from  the smearing
radius 
of the gluons, $N_g$, but in principle they could be set equal.},
the offending diagrams in fig.~\ref{fig:qcdOneLoop} scale as 
\begin{eqnarray}
\delta\Gamma^{(5b,5c),2,3}
& \sim  & \alpha_s\ a^2\ \Lambda_g^2
\ \sim\ {\alpha_s \over N_g^2}
\ \ \ \ ,
\label{eq:42b}
\end{eqnarray}
due to the suppression of the high momentum modes in the gluon propagator.

The natural question to ask here is what is the scale of the coupling
in this process? Note that the bare coupling constant of lattice QCD
suffers from large renormalization as discussed before, so a better-behaved
weak coupling expansion of the lattice quantities uses a renormalized
coupling constant as the expansion parameter. As is suggested by Lepage
and Mackenzie \cite{Lepage}, one first fixes the renormalization
scheme by determining the renormalized coupling $\alpha_{s}^{ren}\left(k^{*}\right)$
from a physical quantity such as the heavy quark potential. Then the
scale of the coupling is set by the typical momentum of the gluon
in a given process. In the case considered above, the energy scale
of the strong coupling constant is dictated by the scale of the
gluon smearing region as the dominant contribution to the integral comes from
this region of the 
integration: $k^{*}\sim\pi/(N_g a)$. A better estimate
of the scale can be obtained by the method explained in Ref.~\cite{Lepage},
but since we are interested in the continuum limit where $a\rightarrow0$,
this is already a reliable estimation of the momentum scale of the
running coupling.

The analysis in QCD 
is more complex at one-loop level than in the scalar theory
due to the presence of the gauge-link required to render the
operator gauge-invariant.  
We have found that the contributions from the operator defined in
eq.~(\ref{eq:25}) scale in the same way as those in the scalar theory,
with the violation of rotational symmetry  suppressed by factors of $\sim
1/N^2$,  but 
both tadpole improvement of the extended links and
smearing of the gauge-field is required.  
Our analysis of Wilson fermions reveals the contributions to matrix elements
that violate rotational invariance in the
continuum limit at the one-loop level are suppressed by factors of 
$\sim \alpha_s/N^2$ and $\sim \alpha_s/N_g^2$, and thus for a smearing defined in physical units,
deviations from rotational invariance scale as  ${\cal O}(a^2)$.  
Contributions that scale as $\sim \alpha_s/N$ and are proportional to the
Wilson parameter, conserve angular momentum and can be absorbed by the operator Z-factor.
Most importantly, as in the scalar theory, there are no mixings with lower dimension operators that
diverge as inverse powers of the lattice spacing.

\section{Summary and Conclusions}

In this paper, a mechanism for the restoration of rotational symmetry
in the continuum limit of lattice field theories is considered. The
essence of this approach is to construct an appropriate operator
on the cubic lattice which has maximum overlap onto the states with
definite angular momentum in the continuum. In analogy to the operator
smearing proposals given in Refs.~\cite{DudekI,DudekII,Edwards} and 
Refs.~\cite{Meyer,JohnsonRW},
the operator is constructed on multiple lattice sites.
Using spherical harmonics in the definition of the operator
is key to having  the leading contributions
to the classical operator be those with the desired angular momentum.
The
sizes of the contributions are controlled by the scale of the smearing of the operator, 
with sub-leading
contributions to both lower and higher dimensional operators
that violate rotational symmetry
being suppressed by $1/N^{2}$ - reflective of the pixelation of the
operator and fields. 
The  $\lambda\phi^{4}$ scalar field theory
is shown to preserve this universal scaling of the leading non-rotationally
invariant contributions at all orders in perturbation theory, compatible
with the finite size scaling results of $\lambda\phi^{4}$-type theories
near their rotational invariant fixed points~\cite{CampostriniI,CampostriniII}.
The same can be shown to be true in $g\phi^3$ scalar field theory.

Gauge invariance somewhat complicates the construction and analysis of analogous
operators in QCD.  Although the tree-level lattice operator in QCD exhibits the
same scaling 
properties as the scalar operator, extended gauge-links connecting the quark fields generate gluonic
interactions that contribute to loop diagrams that are power-law divergent.
Such contributions are either eliminated by tadpole improvement of the extended
links, or are suppressed by smearing of the gauge-field.  
We find that it is the physical length-scales
and continuum renormalization-scale that dictate the size of matrix elements.
The leading non-continuum corrections from the one-loop diagrams preserve
angular momentum, scaling as $\sim \alpha_s a$ for Wilson fermions, and can be
absorbed by the operator $Z$-factor. 
In contrast, contributions that violate rotational symmetry are suppressed by $\alpha_s a^2$ as $a\rightarrow 0$. 
While we have chosen a specific form for the smeared operator, we expect that
the results, in particular the scaling of the violations to rotational symmetry,
are general features of a smeared operator with any (smooth) profile. Also, it is worth mentioning that although the calculations preformed in this work, and the subsequent conclusions, relate operators and matrix elements in H(3) to those in O(3), the methodology and results are expected to hold in relations between H(4) and O(4).  Instead of working with operators formed with spherical harmonics to recover SO(3) invariance, one would work with operators formed with hyper-spherical harmonics to recover O(4) symmetry.

We conclude the paper by discussing the practicality of our result
for the current LQCD calculations as well as its connection to the infra-red
(IR) rotational invariance
recovery of the lattice theories:
\begin{figure}
\begin{centering}
\includegraphics[scale=0.5]{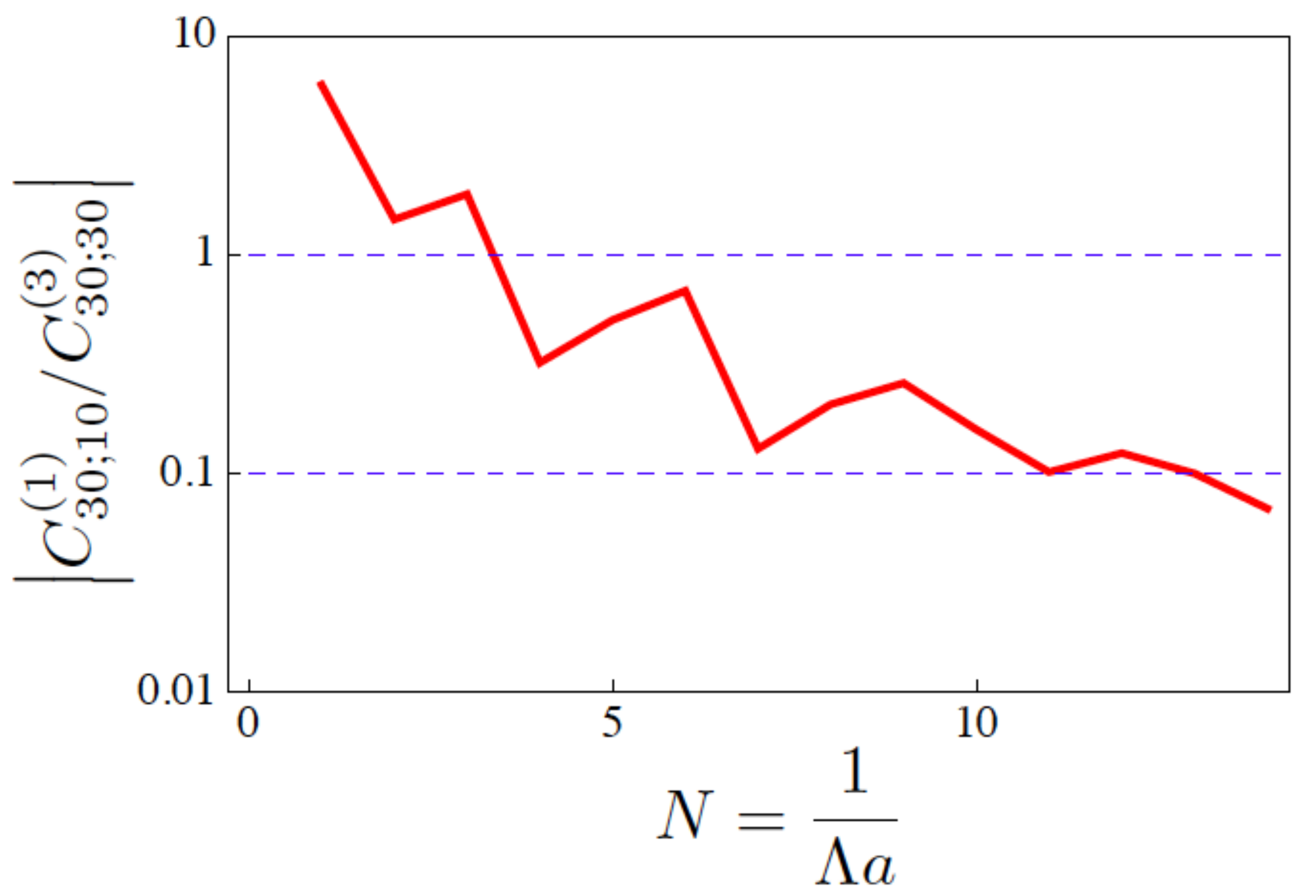}
\par\end{centering}
\caption{{\small 
The absolute value of the ratio of the tree-level coefficient,
    $C_{30;10}^{(1)}$,  of 
a lowest dimension operator  with $L=1$ 
to the tree-level coefficient, $C_{30;30}^{(3)}$, of the lowest dimension operator with
angular momentum, $L=3$, resulting from the $L=3$ operator in eq. (\ref{eq:1}),
as a function of the number of included point-shells.
}}
\label{fig:ratioplot}
\end{figure}
\begin{itemize}
\item
It is important to understand and to quantify 
the violation of angular momentum conservation in the
states and matrix elements calculated using Lattice QCD
with the lattice spacings currently employed. 
One interesting result is 
that by using the tadpole-improved operator extended over several lattice sites
and built from the smeared gauge links, 
the quantum corrections introduce non-continuum corrections to
the tree-level results that are suppressed by at least $\alpha_s$, i.e. they do
not introduce power-divergent contributions.
As an example, 
suppose that a lattice calculation aims to determine a matrix element of an
operator with $L=3$. 
Then, as is demonstrated in fig. (\ref{fig:ratioplot}), 
the coefficient of the lower dimensional derivative operator with $L=1$ is almost $10$
times larger  than the coefficient of the $L=3$ derivative operator 
when the operator is defined  over one lattice site, $N=1$. 
The computational time required to accurately perform the subtraction of the
$L=1$ contribution is significant for  a smearing scale of, say, $\Lambda\sim 2~{\rm GeV}$.
Fortunately, by halving the lattice spacing and smearing the operator over just two point shells ($N=2$), 
the contamination from the lower dimensional operator is reduced by
a factor of $\sim 3$, requiring a factor of $\sim 10$ less computational resources
to accurately perform the subtraction at the same level of precision.
Further, by smearing the operator over ten point shells, 
the contamination from the lower dimensional operator is reduced to 
$\sim 1\%$ of its value at $N=1$.
Given that the lattice spacing associated with  $\Lambda = 2~{\rm GeV}$
is $a\sim 0.1~{\rm fm}$ for $N=1$, to be able to smear out to  the $N=2$ shell requires a
lattice spacing of $a\sim 0.05~{\rm fm}$, pushing the limits of current lattice
generation.  To smear out to the $N=10$ shell would require  a lattice spacing
of  $a\sim 0.01~{\rm fm}$ which is currently impractical.

\item The restoration of rotational invariance as discussed in this paper
regards only the UV asymptote of the lattice theories: as one reaches
a good pixelation of a region of space where the lattice operator
probes, the identification of eigenstates of the angular momentum
operator becomes possible. In the other words, the more point-shells
included in the lattice operator, the larger  overlap the operator
has onto a definite angular momentum state. However, the full recovery
of rotational invariance in the lattice theories requires the suppression
of rotational symmetry breaking contributions to the physical quantities
not only as a result of short-distance discretization effects, but
also as a result of boundary effects of the finite cubic lattice in the
IR regime of the theories. The finite size of the lattice imposes
(anti-)periodic boundary conditions on the lattice wavefunctions which
enforces the lattice momenta to be discretized, 
${\bf p}=\frac{2\pi {\bf n} }{L}$,
where $L$ is the spatial extent of the lattice 
and ${\bf n}$ is a vector of
integers. The IR rotational invariant theory is achieved as the
lattice becomes infinitely large, 
corresponding to a large number of point-shells in the momentum space. 
However, beyond this intuitive picture, one
needs to examine in a quantitative way how this recovery takes place
in the large volume limits of the lattice theories in the same way
as it was discussed for small lattice spacing limit of the theories.
One quantitative explanation of this IR recovery, has been given recently
in Ref.~\cite{Luu} in the context of the 
extraction of phase shifts in higher
partial-waves from the energies of scattering particles in a finite volume
using L\"uschers method.
The idea is that as one includes higher momentum shells, the number
of occurrence (multiplicity) of any given irrep of the cubic group increases. 
As a result, for a fixed energy in the large volume limit,
linear combinations of different states of a given irrep can be formed
which can be shown to be energy eigenstates; and 
the energy-shift of each combination due to interactions is suppressed 
in all but one partial-wave in the infinite-volume limit. 
So, although each irrep state has an
overlap onto infinitely many angular momentum states, the high multiplicity
of a given irrep in a large momentum shell generates energy-eigenstates
which 
dominantly overlap onto states of definite angular momentum,
and the mixing with other angular momentum states becomes insignificant in the
large volume limit.
This picture also helps to better understand the mechanism of the
UV rotational invariance recovery due to the operator smearing. 
It
is the high multiplicity of the irreps in large (position-space) 
shells that  is responsible
for projecting out a definite angular momentum eigenstate. 
These large
shells are obtained by reducing the pixelation of the lattice by taking
$a\rightarrow 0$ in position-space,  or increasing the size of the
lattice by taking $L\rightarrow\infty$ in momentum-space - both are required in
order to recover rotational invariance from calculations performed on a lattice.
\end{itemize}

\subsection*{Acknowledgment}

We would like to thank David B. Kaplan and Sichun Sun for their contributions
to a precursor to this work, Saul Cohen, David B. Kaplan, Kostas
Orginos, Stephen R. Sharpe and Boris Spivak for useful discussions, and David B. Kaplan and Thomas C. Luu for their feedback on the earlier draft of this paper. ZD and MJS were supported
in part by the DOE grant DE-FG03-97ER4014.

\appendix


\section{Operator Basis}
\label{app:operators}
\noindent
In this appendix, a basis for composite local operators is presented. 
Any local operator that is  bilinear in the scalar field 
with $L$ spatial indices, and that is invariant under cubic transformations,
can be written as
\begin{equation}
\mathcal{O}_{i_{1}i_{2}...i_{L}}^{\left(d\right)}\left(\mathbf{x}\right)
\ =\ 
\phi^{\dagger}\left(\mathbf{x}\right)
\ Q_{i_{1}i_{2}...i_{L}}^{\left(d\right)}
\ \phi\left(\mathbf{x}\right)
 \ \ \ ,
\label{eq:45}
\end{equation}
where $Q_{i_{1}i_{2}...i_{L}}^{\left(d\right)}$ is a homogeneous
function of the operator ${\nabla}_{i}$, 
and degree $d$ ($d\geq L$) is defined to be the number of ${\nabla}$'s. 
Their forms are determined by the symmetric traceless tensor of rank $L$ 
that  respect cubic symmetry constructed from $d$ ${\nabla}$'s.
The operators composed of fewer than seven derivatives and 
with no spatial indices  
are
\begin{eqnarray}
\mathcal{O}^{\left(0\right)}\left(\mathbf{x}\right)
& = & 
\phi^{\dagger}\left(\mathbf{x}\right)\phi\left(\mathbf{x}\right)
\nonumber\\
\mathcal{O}^{\left(2\right)}\left(\mathbf{x}\right)
&  = & 
\phi^{\dagger}\left(\mathbf{x}\right)\mathbf{\nabla}^{2}\phi\left(\mathbf{x}\right)
\nonumber\\
\mathcal{O}^{\left(4\right)}\left(\mathbf{x}\right)
& = & 
\phi^{\dagger}\left(\mathbf{x}\right)\left(\mathbf{\nabla}^{2}\right)^{2}\phi\left(\mathbf{x}\right)
\nonumber\\
\mathcal{O}^{\left(4,RV\right)}\left(\mathbf{x}\right)
& = &
\phi^{\dagger}\left(\mathbf{x}\right)\ \sum_{j}\nabla_{j}^{4}\ \phi\left(\mathbf{x}\right)
\nonumber\\
\mathcal{O}^{\left(6\right)}\left(\mathbf{x}\right)
& = & 
\phi^{\dagger}\left(\mathbf{x}\right)\left(\mathbf{\nabla}^{2}\right)^{3}\phi\left(\mathbf{x}\right)
\nonumber\\
\mathcal{O}^{\left(6,RV;1\right)}\left(\mathbf{x}\right)
& = & 
\phi^{\dagger}\left(\mathbf{x}\right) \mathbf{\nabla}^{2} \
\sum_{j}\nabla_{j}^{4}\ \phi\left(\mathbf{x}\right)
\nonumber\\
\mathcal{O}^{\left(6,RV;2\right)}\left(\mathbf{x}\right)
& = &
\phi^{\dagger}\left(\mathbf{x}\right)\ \sum_{j}\nabla_{j}^{6}\ \phi\left(\mathbf{x}\right)
\ \ \ \ .
\label{eq:46}
\end{eqnarray}
Except for three of these operators which explicitly
break the rotational symmetry, 
they transform as $L=0$ under rotations.

The operators with one spatial index with up to six derivatives
are
\begin{eqnarray}
\mathcal{O}_{i}^{\left(1\right)}\left(\mathbf{x}\right)
& = & 
\phi^{\dagger}\left(\mathbf{x}\right)\mathbf{\nabla}_{i}\phi\left(\mathbf{x}\right)
\nonumber\\
\mathcal{O}_{i}^{\left(3\right)}\left(\mathbf{x}\right)
& = &
\phi^{\dagger}\left(\mathbf{x}\right)\mathbf{\nabla}^{2}\mathbf{\nabla}_{i}\phi\left(\mathbf{x}\right)
\nonumber\\
\mathcal{O}_{i}^{\left(5\right)}\left(\mathbf{x}\right)
 & = & 
\phi^{\dagger}\left(\mathbf{x}\right)\left(\mathbf{\nabla}^{2}\right)^{2}\mathbf{\nabla}_{i}\phi\left(\mathbf{x}\right)
\nonumber\\
\mathcal{O}_{i}^{\left(5,RV\right)}\left(\mathbf{x}\right)
 & = & \phi^{\dagger}\left(\mathbf{x}\right)\sum_{j}\nabla_{j}^{4}\mathbf{\nabla}_{i}\phi\left(\mathbf{x}\right)
 \ \ .
\label{eq:47}
\end{eqnarray}
There is one operator which breaks rotational invariance, and
the rest  transform as $L=1$ under rotations.

The operators with two  spatial index with up to six derivatives
are
\begin{eqnarray}
\mathcal{O}_{ij}^{\left(2\right)}\left(\mathbf{x}\right)
 & = & 
\phi^{\dagger}\left(\mathbf{x}\right)
\left[\mathbf{\nabla}_{i}\mathbf{\nabla}_{j}-\frac{1}{3}\delta_{ij}\mathbf{\nabla}^{2}\right]
\phi\left(\mathbf{x}\right)
\nonumber\\
\mathcal{O}_{ij}^{\left(4\right)}\left(\mathbf{x}\right)
& = &
\phi^{\dagger}\left(\mathbf{x}\right)
\mathbf{\nabla}^{2}\left[\mathbf{\nabla}_{i}\mathbf{\nabla}_{j}-\frac{1}{3}\delta_{ij}\mathbf{\nabla}^{2}\right]
\phi\left(\mathbf{x}\right)
\nonumber\\
\mathcal{O}_{ij}^{\left(6\right)}\left(\mathbf{x}\right)
& = &
\phi^{\dagger}\left(\mathbf{x}\right)
\left(\mathbf{\nabla}^{2}\right)^{2}\left[\mathbf{\nabla}_{i}\mathbf{\nabla}_{j}
-\frac{1}{3}\delta_{ij}\mathbf{\nabla}^{2}\right]
\phi\left(\mathbf{x}\right)
\nonumber\\
\mathcal{O}_{ij}^{\left(6,RV\right)}\left(\mathbf{x}\right)
 & = & 
\phi^{\dagger}\left(\mathbf{x}\right)
\sum_{k}\nabla_{k}^{4}\left[\mathbf{\nabla}_{i}\mathbf{\nabla}_{j}
-\frac{1}{3}\delta_{ij}\mathbf{\nabla}^{2}\right]
\phi\left(\mathbf{x}\right)
\ \ \ .
\label{eq:48}
\end{eqnarray}
There is one operator which breaks rotational invariance, and
the rest  transform as $L=2$ under rotations.

Operators with three, four and five spatial indices which have $L=3$,
$L=4$ and $L=5$ respectively are listed below. There is no operator
which breaks rotational invariance up to six derivatives:

\begin{eqnarray}
\mathcal{O}_{ijk}^{\left(3\right)}\left(\mathbf{x}\right)
 & = & 
\phi^{\dagger}\left(\mathbf{x}\right)
\left[\mathbf{\nabla}_{i}\mathbf{\nabla}_{j}\mathbf{\nabla}_{k}
-\frac{1}{5}\mathbf{\nabla}^{2}
\left(\delta_{ij}\mathbf{\nabla}_{k}+\delta_{jk}\mathbf{\nabla}_{i}+\delta_{ki}\mathbf{\nabla}_{j}\right)\right]
\phi\left(\mathbf{x}\right)
\nonumber\\
\mathcal{O}_{ijk}^{\left(5\right)}\left(\mathbf{x}\right)
& = & 
\phi^{\dagger}\left(\mathbf{x}\right)
\mathbf{\nabla}^{2}\left[\mathbf{\nabla}_{i}\mathbf{\nabla}_{j}\mathbf{\nabla}_{k}
-\frac{1}{5}\mathbf{\nabla}^{2}
\left(\delta_{ij}\mathbf{\nabla}_{k}+\delta_{jk}\mathbf{\nabla}_{i}+\delta_{ki}\mathbf{\nabla}_{j}\right)\right]
\phi\left(\mathbf{x}\right)
\ \ \ ,
\label{eq:49}
\end{eqnarray}
\begin{eqnarray}
\mathcal{O}_{ijkl}^{\left(4\right)}\left(\mathbf{x}\right)
& = &  
\phi^{\dagger}\left(\mathbf{x}\right)
\left[
\mathbf{\nabla}_{i}\mathbf{\nabla}_{j}\mathbf{\nabla}_{k}\mathbf{\nabla}_{l}
\phantom{\frac{1}{7}}
\right.\nonumber\\
&&\left.
\qquad
\ -\
\frac{1}{7}\mathbf{\nabla}^{2}\left(\delta_{ij}\mathbf{\nabla}_{k}\mathbf{\nabla}_{l}
+\delta_{ik}\mathbf{\nabla}_{j}\mathbf{\nabla}_{l}
+\delta_{il}\mathbf{\nabla}_{k}\mathbf{\nabla}_{j}+\delta_{jk}\mathbf{\nabla}_{i}\mathbf{\nabla}_{l}
+\delta_{jl}\mathbf{\nabla}_{i}\mathbf{\nabla}_{k}+\delta_{kl}\mathbf{\nabla}_{i}\mathbf{\nabla}_{j}\right)
\right.\nonumber\\
&&\left.
\qquad
\ +\
\frac{1}{35}\left(\mathbf{\nabla}^{2}\right)^{2}\left(\delta_{ij}\delta_{kl}
+\delta_{ik}\delta_{jl}+\delta_{il}\delta_{jk}\right)
\right]
\phi\left(\mathbf{x}\right)
\nonumber\\
\mathcal{O}_{ijkl}^{\left(6\right)}\left(\mathbf{x}\right)
& = & 
\phi^{\dagger}\left(\mathbf{x}\right)
\mathbf{\nabla}^{2}
\left[
\mathbf{\nabla}_{i}\mathbf{\nabla}_{j}\mathbf{\nabla}_{k}\mathbf{\nabla}_{l}
\phantom{\frac{1}{7}}
\right.\nonumber\\
&&\left.
\qquad
\ -\
\frac{1}{7}\mathbf{\nabla}^{2}\left(\delta_{ij}\mathbf{\nabla}_{k}\mathbf{\nabla}_{l}
+\delta_{ik}\mathbf{\nabla}_{j}\mathbf{\nabla}_{l}+\delta_{il}\mathbf{\nabla}_{k}\mathbf{\nabla}_{j}
+\delta_{jk}\mathbf{\nabla}_{i}\mathbf{\nabla}_{l}+\delta_{jl}\mathbf{\nabla}_{i}\mathbf{\nabla}_{k}
+\delta_{kl}\mathbf{\nabla}_{i}\mathbf{\nabla}_{j}\right)
\right. \nonumber\\
&& \left.
\qquad
\ +\ 
\frac{1}{35}\left(\mathbf{\nabla}^{2}\right)^{2}\left(\delta_{ij}\delta_{kl}+\delta_{ik}\delta_{jl}
+\delta_{il}\delta_{jk}\right)
\right]
\phi\left(\mathbf{x}\right)
\ \ \ ,
\label{eq:50}
\end{eqnarray}

\begin{eqnarray}
\mathcal{O}_{ijklm}^{\left(5\right)}\left(\mathbf{x}\right)
& = & \phi^{\dagger}\left(\mathbf{x}\right)
\left[
\mathbf{\nabla}_{i}\mathbf{\nabla}_{j}\mathbf{\nabla}_{k}\mathbf{\nabla}_{l}\mathbf{\nabla}_{m}
\right.
\nonumber\\
&& \left.
-\frac{1}{7}\mathbf{\nabla}^{2}
\left(
\delta_{ij}\mathbf{\nabla}_{k}\mathbf{\nabla}_{l}\mathbf{\nabla}_{m}
+\delta_{ik}\mathbf{\nabla}_{j}\mathbf{\nabla}_{l}\mathbf{\nabla}_{m}
+\delta_{il}\mathbf{\nabla}_{k}\mathbf{\nabla}_{j}\mathbf{\nabla}_{m}
+\delta_{im}\mathbf{\nabla}_{k}\mathbf{\nabla}_{l}\mathbf{\nabla}_{j}
\right.\right.
\nonumber\\
&& \left.
\qquad\qquad
+\delta_{jk}\mathbf{\nabla}_{i}\mathbf{\nabla}_{l}\mathbf{\nabla}_{m}
+\delta_{jl}\mathbf{\nabla}_{k}\mathbf{\nabla}_{i}\mathbf{\nabla}_{m}
+\delta_{jm}\mathbf{\nabla}_{k}\mathbf{\nabla}_{i}\mathbf{\nabla}_{l}
+\delta_{kl}\mathbf{\nabla}_{i}\mathbf{\nabla}_{j}\mathbf{\nabla}_{m}
\right.\nonumber\\
&&\left.
\qquad\qquad
+\delta_{km}\mathbf{\nabla}_{i}\mathbf{\nabla}_{j}\mathbf{\nabla}_{l}
+\delta_{lm}\mathbf{\nabla}_{i}\mathbf{\nabla}_{j}\mathbf{\nabla}_{k}
\right)
\nonumber\\
&& 
\ +\ \frac{1}{63}
\left(\mathbf{\nabla}^{2}\right)^{2}
\left[
\left(
\delta_{ij}\delta_{kl}+\delta_{ik}\delta_{jl}+\delta_{il}\delta_{jk}
\right)\mathbf{\nabla}_{m}
+
\left(
\delta_{ij}\delta_{km}+\delta_{ik}\delta_{jm}+\delta_{im}\delta_{jk}\right)\mathbf{\nabla}_{l}
\right.\nonumber\\
&&
\left.
\qquad\qquad
+
\left(\delta_{ij}\delta_{ml}+\delta_{im}\delta_{jl}+\delta_{il}\delta_{jm}\right)\mathbf{\nabla}_{k}
+
\left(\delta_{im}\delta_{kl}+\delta_{ik}\delta_{ml}+\delta_{il}\delta_{mk}\right)\mathbf{\nabla}_{j}
\right.\nonumber\\
&&
\left.\left.
\qquad\qquad
+
\left(\delta_{mj}\delta_{kl}+\delta_{mk}\delta_{jl}+\delta_{ml}\delta_{jk}\right)\mathbf{\nabla}_{i}
\right]
\right]
\phi\left(\mathbf{x}\right)
\ \ \ .
\label{eq:51}
\end{eqnarray}

Note that as demonstrated in eq.~(\ref{eq:46}), there can be more than one operator
that breaks rotational invariance at a given order
in derivative expansion. 
To arrive at a notation that is general and useful,
one can use the fact that any cubically invariant polynomial of a 
three-vector $\mathbf{V}$, can be expanded in terms of only three  cubically
invariant structures,
\begin{equation}
\sum_{k}V_{k}^{2}\ ,\ 
\sum_{k}V_{k}^{4}\ ,\ 
\sum_{k}V_{k}^{6}
\ \ \ .
\label{eq:52}
\end{equation}
The number of times each structure appears in a derivative
operator, as well as the number of free indices,
uniquely specify the operator.
For example,  with nine derivatives and one spatial
index, one can make four independent   operators,
\begin{eqnarray}
\mathcal{O}_{i}^{\left(4,0,0\right)}\left(\mathbf{x}\right)
 & = & 
\phi^{\dagger}\left(\mathbf{x}\right)\left(\mathbf{\nabla}^{2}\right)^{4}\mathbf{\nabla}_{i}\phi\left(\mathbf{x}\right)
\nonumber\\
\mathcal{O}_{i}^{\left(2,1,0\right)}\left(\mathbf{x}\right)
& = & 
\phi^{\dagger}\left(\mathbf{x}\right)\left(\mathbf{\nabla}^{2}\right)^{2}
\left(\sum_{k}\nabla_{k}^{4}\right)\mathbf{\nabla}_{i}\phi\left(\mathbf{x}\right)
\nonumber\\
\mathcal{O}_{i}^{\left(1,0,1\right)}\left(\mathbf{x}\right)
& = &
\phi^{\dagger}\left(\mathbf{x}\right)\left(\mathbf{\nabla}^{2}\right)\left(\sum_{j}\nabla_{j}^{6}\right)\mathbf{\nabla}_{i}\phi\left(\mathbf{x}\right)
\nonumber\\
\mathcal{O}_{i}^{\left(0,2,0\right)}\left(\mathbf{x}\right)
& = & 
\phi^{\dagger}\left(\mathbf{x}\right)\left(\sum_{k}\nabla_{k}^{4}\right)^{2}\mathbf{\nabla}_{i}\phi\left(\mathbf{x}\right)
\ \ \ ,
\label{eq:53}
\end{eqnarray}
and generally,
\begin{eqnarray}
\mathcal{O}_{i}^{\left(m,n,p\right)}\left(\mathbf{x}\right)
& = & 
\left(\mathbf{\nabla}^{2}\right)^{m}
\left(\sum_{k}\nabla_{k}^{4}\right)^n
\left(\sum_{k}\nabla_{k}^{6}\right)^p
\mathbf{\nabla}_{i}
\phi\left(\mathbf{x}\right)
\ \ \ .
\label{eq:53b}
\end{eqnarray}
It is then obvious that $d=2m+4n+6p+L$
gives the total number of derivatives in the operator, where $L$ is
the number of free indices. For $n=p=0$, the operator is rotationally
invariant with angular momentum $L$.

\section{Rotational Invariance Violating Coefficients : An Example}
\label{app:RIviolation}

In this appendix, an explicit derivation of a rotational invariance
violating coefficient in both coordinate-space, and momentum-space
formalism, introduced in section~\ref{sec:Classical}, 
is presented.
Consider the position space operator 
$\hat{\theta}_{00}^{\left(4\right)}\left(\mathbf{x};a,N\right)$
where superscript indicates that only operators with
four derivatives are retained in the expansion of $\hat{\theta}_{00}$.
The goal is to derive the LO correction to the continuum
values of coefficients $C_{00,00}^{\left(4\right)}$ 
and $C_{00,00}^{\left(4;RV\right)}$:
\begin{eqnarray}
\hat{\theta}_{00}^{\left(4\right)}\left(\mathbf{x};a,N\right)
& = & 
\phi\left(\mathbf{x}\right)\left[\left(Na\right)^{4}C_{00,00}^{\left(4\right)}\left(\nabla^{2}\right)^{2}
+\left(Na\right)^{4}C_{00,00}^{\left(4;RV\right)}
\left(\nabla_{x}^{4}+\nabla_{y}^{4}+\nabla_{z}^{4}\right)\right]
\phi\left(\mathbf{x}\right)
\nonumber\\
& = & 
\frac{3}{4\pi}\frac{\left(aN\right)^{4}}{4!}
\ \sum_{\mathbf{P}}
\ \int_{0}^{1}dy\ y^{6}\ \int
d\Omega_{\mathbf{y}}\ 
e^{i2\pi  N\mathbf{p}\cdot\mathbf{y}}
\phi^{\dagger}\left(\mathbf{x}\right)
\left(\hat{\mathbf{y}} \cdot \mathbf{\nabla}\right)^{4}
\phi\left(\mathbf{x}\right)Y_{00}\left(\Omega_{\mathbf{y}}
\right)
\ \ \ .
\qquad
\label{eq:54}
\end{eqnarray}
The $y$ integration is
\begin{eqnarray}
\int_{0}^{1}dy\ y^{6}\ 
&& \int d\Omega_{\mathbf{y}}\ e^{i2\pi N\mathbf{p}\cdot\mathbf{y}}y^{i}y^{j}y^{k}y^{l}
\ =\ 
\alpha\left(p^{i}p^{j}p^{k}p^{l}\right) 
\ +\ \gamma\left(\delta^{ij}\delta^{kl}+\delta^{ik}\delta^{jl}+\delta^{il}\delta^{jk}\right)
\nonumber\\
& & +\beta\left(p^{i}p^{j}\delta^{kl}+p^{i}p^{k}\delta^{jl}
+p^{i}p^{l}\delta^{jk}+p^{k}p^{l}\delta^{ij}+p^{j}p^{l}\delta^{ik}+p^{j}p^{k}\delta^{il}
\right)
\ \ \ ,
\label{eq:55}
\end{eqnarray}
and the coefficients $\alpha$, $\beta$ and $\gamma$ can be determined.
It is easy to see that coefficient $\alpha$ 
makes the dominant contribution 
in the large $N$ limit. 
Using
\begin{equation}
\sum_{\mathbf{p}}f\left(p^{2}\right)\left(\mathbf{p}.\mathbf{A}\right)^{4}=\sum_{\mathbf{p}}f\left(p^{2}\right)\left(\rho\left|\mathbf{A}\right|^{4}+\sigma\sum_{j}\left(A^{j}\right)^{4}\right)
\ \ \ ,
\label{eq:56}
\end{equation}
for any rotational invariant function $f$ of the vector $\mathbf{p}$,
with 
\begin{eqnarray}
\rho & = & 
\frac{1}{2}\left(\left|\mathbf{p}\right|^{4}-3p_{z}^{4}\right)
\ \ ,\ \ 
\sigma\ =\ 
\frac{1}{2}\left(5p_{z}^{4}-\left|\mathbf{p}\right|^{4}\right)
\ \ \ ,
\label{eq:sigrho}
\end{eqnarray}
one finds that the deviations of $C_{00,00}^{\left(4\right)}$
and $C_{00,00}^{\left(4;RV\right)}$ from their continuum values are
\begin{eqnarray}
\delta
C_{00,00}^{\left(4\right)}
& = & 
\frac{1}{96\sqrt{\pi}}
\ \sum_{\mathbf{p}\neq\mathbf{0}}
\left(-\frac{3\cos\left(2\pi
      N\left|\mathbf{p}\right|\right)}{4\pi^{2}
\left|\mathbf{p}\right|^{6}N^{2}}\right)\left(-3p_{z}^{4}+\left|\mathbf{p}\right|^{4}\right)
\nonumber\\
\delta C_{00,00}^{\left(4;RV\right)}
& = & \frac{1}{96\sqrt{\pi}}\sum_{\mathbf{p}\neq\mathbf{0}}
\left(-\frac{3\cos\left(2\pi N\left|\mathbf{p}\right|\right)}{4\pi^{2}
\left|\mathbf{p}\right|^{6}N^{2}}\right)\left(5p_{z}^{4}-\left|\mathbf{p}\right|^{4}\right)
\ \ \ .
\label{eq:57}
\end{eqnarray}

The emergence of rotational invariance violating coefficients from
the momentum-space construction  is somewhat less obvious. 
From eq.~(\ref{eq:14}) and eq.~(\ref{eq:15})
the operator $\hat{\tilde{\theta}}_{00}^{\left(4\right)}\left(\mathbf{k};a,N\right)$
can be written as
\begin{eqnarray}
\hat{\tilde{\theta}}_{00}^{\left(4\right)}\left(\mathbf{k};a,N\right)
& = & 
\tilde{\phi}\left(\mathbf{k}\right)
\tilde{\phi}\left(-\mathbf{k}\right)
\left[\left(Na\right)^{4}C_{00,00}^{\left(4\right)}\left|\mathbf{k}\right|^{4}
+\left(Na\right)^{4}C_{00,00}^{\left(4;RV\right)}(k_{x}^{4}+k_{y}^{4}+k_{z}^{4})\right]
\nonumber\\
& = & \tilde{\phi}\left(\mathbf{k}\right)
\tilde{\phi}\left(-\mathbf{k}\right)
6\sqrt{\pi}\ 
\sum_{\mathbf{p}}\sum_{L_{1},M_{1},L_{2},M_{2}}i^{L_{1}+L_{2}}\sqrt{\frac{\left(2L_{1}+1\right)
\left(2L_{2}+1\right)}{2L+1}}
\nonumber\\
&&
\times
\left\langle L_{1}0;L_{2}0\left|00\right.\right\rangle 
\left\langle L_{1}M_{1};L_{2}M_{2}\left|00\right.\right\rangle
\
Y_{L_{1}M_{1}}\left(\Omega_{\hat{k}}\right)Y_{L_{2}M_{2}}\left(\Omega_{\hat{p}}\right)
\nonumber\\
&&
\times
 \int_{0}^{1}dy\ y^{2}\ j_{L_{1}}\left(aN\left|\mathbf{k}\right|y\right)
\ j_{L_{2}}\left(2\pi
  N\left|\mathbf{p}\right|y\right)
\Bigl\lvert_{k^4}
\ \ \ ,
\label{eq:58}
\end{eqnarray}
where only the terms of order $k^4$ are retained from the integral.
As such,
only $L_1=4$ with 
$Y_{4\pm4}\left(\Omega_{\hat{p}}\right)$ and
$Y_{40}\left(\Omega_{\hat{p}}\right)$, and 
$L_1=0$ with $Y_{00}\left(\Omega_{\hat{p}}\right)$, contribute to the sum.
This reduces the relation to 
\begin{eqnarray}
\hat{\tilde{\theta}}_{00}^{\left(4\right)}\left(\mathbf{k};a,N\right)
& = & 
6\sqrt{\pi}\ 
\sum_{\mathbf{p}}\left\{
\   Y_{00}\left(\Omega_{\hat{k}}\right)Y_{00}\left(\Omega_{\hat{p}}\right)
\ \int_{0}^{1}dy\ y^{2}\ 
\frac{\left(aN\left|\mathbf{k}\right|y\right)^{4}}{120}j_{0}\left(2\pi N\left|\mathbf{p}\right|y\right)
\right.
\nonumber\\
& + &
9\left[
\left\langle 40;40\left|00\right.\right\rangle ^{2}
Y_{40}\left(\Omega_{\hat{k}}\right)Y_{40}
\left(\Omega_{\hat{p}}\right)+\left\langle 40;40\left|00\right.\right\rangle 
\left\langle 44;4-4\left|00\right.\right\rangle
Y_{44}\left(\Omega_{\hat{k}}\right)
Y_{4-4}\left(\Omega_{\hat{p}}\right)\right.
\nonumber\\
&&
\left.\left.+\left\langle 40;40\left|00\right.\right\rangle \left\langle
      4-4;44\left|00\right.\right\rangle
    Y_{4-4}\left(\Omega_{\hat{k}}\right)Y_{44}\left(\Omega_{\hat{p}}\right)
\right]
\right.
\nonumber\\
& & \left. 
\qquad\qquad
\times
\int_{0}^{1}dy\ y^{2}\ 
\frac{\left(aN\left|\mathbf{k}\right|y\right)^{4}}{945}
\ j_{4}\left(2\pi N\left|\mathbf{p}\right|y\right)
\ \right\} 
\ \ \ .
\label{eq:60}
\end{eqnarray}
Using the relations
\begin{eqnarray}
\sum_{\mathbf{p}}f\left(p^{2}\right)\left(\left|\mathbf{p}\right|^{4}
Y_{40}\left(\Omega_{\hat{p}}\right)\right)
& = & 
\frac{21}{16}\sqrt{\frac{1}{\pi}}\sum_{\mathbf{p}}f\left(p^{2}\right)
\left(5p_{z}^{4}-\left|\mathbf{p}\right|^{4}\right)
\ \ \ ,
\nonumber\\
\sum_{\mathbf{p}}f\left(p^{2}\right)\left(\left|\mathbf{p}\right|^{4} 
Y_{4\pm4}\left(\Omega_{\hat{p}}\right)\right)
& = & 
\frac{3}{16}\sqrt{\frac{35}{2\pi}}\sum_{\mathbf{p}}f\left(p^{2}\right)
\left(5p_{z}^{4}-\left|\mathbf{p}\right|^{4}\right)
\ \ \ ,
\label{eq:62}
\end{eqnarray}
and keeping the LO term in $1/N$ from the y integration
gives
\begin{eqnarray}
&& \hat{\tilde{\theta}}_{00}^{\left(4\right)}\left(\mathbf{k};a,N\right)
\ = \ 
3\left(aN\left|\mathbf{k}\right|\right)^{4}
\sum_{\mathbf{p}\neq\mathbf{0}}\left(-\frac{\cos\left(2\pi
      N\left|\mathbf{p}\right|\right)}{4\pi\left|\mathbf{p}\right|^{2}N^{2}}\right)\left\{ \frac{1}{120}Y_{00}\left(\Omega_{\hat{k}}\right)
\right.
\nonumber\\
&& \left.
\qquad
+\frac{\sqrt{4\pi}}{945}\left(5p_{z}^{4}-\left|\mathbf{p}\right|^{4}\right)
\left[\frac{21}{16}\sqrt{\frac{1}{\pi}}
Y_{40}\left(\Omega_{\hat{k}}\right)+\frac{3}{16}
\sqrt{\frac{35}{2\pi}}\left(Y_{4-4}
\left(\Omega_{\hat{k}}\right)
+Y_{44}\left(\Omega_{\hat{k}}\right)\right)\right]\right\} 
\ .
\label{eq:63}
\end{eqnarray}
Finally, we use the relation
\begin{equation}
\frac{k_{x}^{4}+k_{y}^{4}+k_{z}^{4}}{\left|\mathbf{k}\right|^{4}}
\ =\ 
\frac{6\sqrt{\pi}}{5}Y_{00}\left(\Omega_{\hat{k}}\right)
+\frac{4\sqrt{\pi}}{15}Y_{40}\left(\Omega_{\hat{k}}\right)
+\frac{2}{3}\sqrt{\frac{2\pi}{35}}\left(Y_{4-4}\left(\Omega_{\hat{k}}\right)
+Y_{44}\left(\Omega_{\hat{k}}\right)\right)
\ \ \ ,
\label{eq:64}
\end{equation}
to identify the
coefficients $\delta C_{00,00}^{\left(4\right)}$ and $\delta C_{00,00}^{\left(4;RV\right)}$
from eq.~(\ref{eq:63})
\begin{eqnarray}
\delta C_{00,00}^{\left(4\right)}
& = & 
\frac{1}{96\sqrt{\pi}}\sum_{\mathbf{p}\neq\mathbf{0}}
\left(-\frac{3\cos\left(2\pi
      N\left|\mathbf{p}\right|\right)}{4\pi^{2}\left|\mathbf{p}\right|^{6}N^{2}}\right)
\left(-3p_{z}^{4}+\left|\mathbf{p}\right|^{4}\right)
\ \ ,
\nonumber\\
\delta C_{00,00}^{\left(4;RV\right)}
& = & \frac{1}{96\sqrt{\pi}}\sum_{\mathbf{p}\neq\mathbf{0}}
\left(-\frac{3\cos\left(2\pi N\left|\mathbf{p}\right|\right)}{4\pi^{2}
\left|\mathbf{p}\right|^{6}N^{2}}\right)\left(5p_{z}^{4}-\left|\mathbf{p}\right|^{4}
\right)
\ \ \ ,
\label{eq:65}
\end{eqnarray}
which recovers the position-space results given in eq.~(\ref{eq:57}).

\section{Matrix Elements for Non-Zero External Momentum}
\label{app:nonzerop}

\begin{figure}[!ht]
\begin{centering}
\includegraphics[scale=0.13]{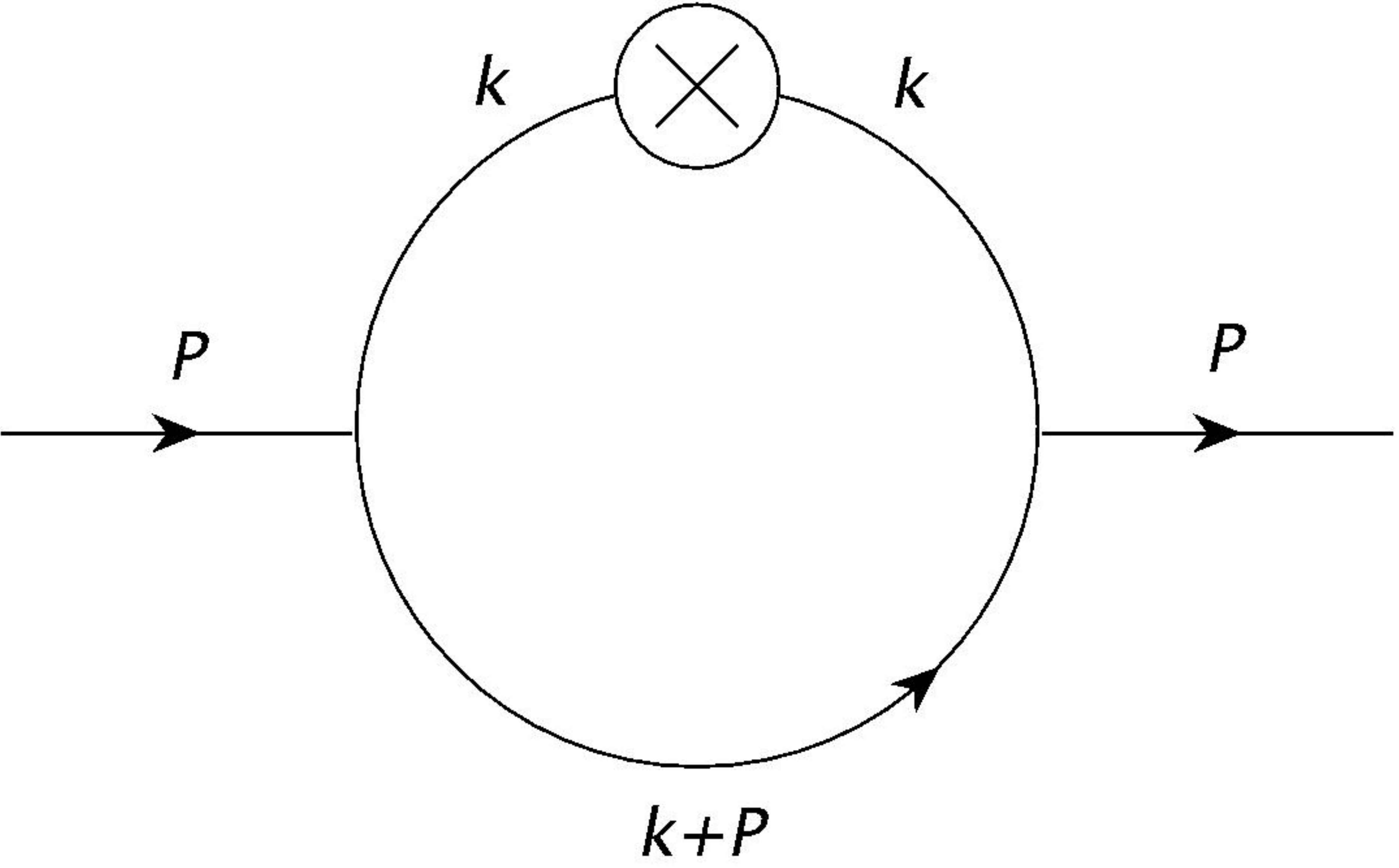}
\par\end{centering}
\caption{{\small One-loop contribution to the two-point function with an insertion
of the operator in $g\phi^{3}$}}
\label{fig:gf3}
\end{figure}

The loop calculations presented in the body of this paper 
have  been performed for vanishing 
external momentum, therefore only the quantum corrections to the $L=0$ operator
have been considered. 
In this appendix, the generalization to non-zero external momentum is
presented,  where the one-loop correction to the two-point function
with an  insertion of the
smeared operator is considered in scalar $g\phi^{3}$ theory, see fig.~\ref{fig:gf3}. 
The loop integral to be evaluated is
\begin{equation}
J_{LM}
\ =\ 
\frac{3}{4\pi
  N^{3}}\sum_{\bf n}^{|{\bf n}|\le N}
\int_{-\frac{\pi}{a}}^{\frac{\pi}{a}}\frac{d^{4}k}{\left(2\pi\right)^{4}}
\frac{e^{i\mathbf{k}\cdot\mathbf{n}a}}{\left(\hat{k}^{2}+m^{2}\right)^{2}
\left(\left(\widehat{k+P}\right)^{2}+m^{2}\right)}
\ Y_{LM}\left(\Omega_{\mathbf{n}}\right)
\ \ \ ,
\label{eq:66}
\end{equation}
where 
\begin{eqnarray}
\hat{k}^{2}& = & 
{4\over a^2}\ \sum_{\mu}
\sin^{2}\left(k_{\mu}a\over 2\right)
\ \ \ ,\ \ \ 
\left(\widehat{k+P}\right)^{2}
\ =\ 
{4\over a^2}\ \sum_{\mu}
\sin^{2}\left(\left(k_{\mu}+P_{\mu}\right)a\over 2\right)
\ \ \ .
\label{eq:momext}
\end{eqnarray}
Note that the operator is smeared over a physical region whose size
is small compared to the hadronic scale, and as a result the external momenta are small compared to
the the scale of the operator $\Lambda=1/Na$. 
Therefore one may perform a Taylor expansion of the loop integral in $P_{i}/\Lambda$ to obtain
\begin{eqnarray}
J_{LM} 
& = & 
\frac{3}{16\pi^4}i^{L}\frac{1}{\Lambda^{2}}
\ \int_{-\pi N}^{\pi N}\ dq_{4}\ d^{3}q
\ \left[\int_{0}^{1}dy\ y^{2}\ 
j_{L}\left(qy\right)\right]
\ Y_{LM}\left(\Omega_{\mathbf{q}}\right)
\nonumber\\
&&
\times
\left(4N^{2}\sum_{i=1}^{3}\sin^{2}\left(\frac{q_{i}}{2N}\right)+4N^{2}\sin^{2}
\left(\frac{q_{4}}{2N}\right)+\frac{m^{2}}{\Lambda^{2}}\right)^{-2}
\nonumber\\
&&
\times
\left(4N^{2}\sum_{i=1}^{3}\sin^{2}\left(\frac{q_{i}}{2N}\right)+4N^{2}\sin^{2}
\left(\frac{q_{4}}{2N}+\frac{P_{4}}{2N\Lambda}\right)+\frac{m^{2}}{\Lambda^{2}}\right)^{-1}
\nonumber\\
&&
\times
\sum_{k=0}^{\infty}\left[-\frac{4N^{2}\sum_{i=1}^{3}\frac{1}{2}\sin\left(\frac{q_{i}}{N}\right)
\sin\left(\frac{P_{i}}{N\Lambda}\right)+4N^{2}\sum_{i=1}^{3}\cos\left(\frac{q_{i}}{N}\right)
\sin^{2}\left(\frac{P_{i}}{2N\Lambda}\right)}{4N^{2}\sum_{i=1}^{3}\sin^{2}
\left(\frac{q_{i}}{2N}\right)+4N^{2}\sin^{2}\left(\frac{q_{4}}{2N}+\frac{P_{4}}{2N\Lambda}\right)
+\frac{m^{2}}{\Lambda^{2}}}\right]^{k}
\ ,
\label{eq:67}
\end{eqnarray}
where $\mathbf{q}=\mathbf{k}/\Lambda$, $q_{4}=k_{4}/\Lambda$, and
only the leading term in the Poisson sum is retained.
As was shown before, the non-zero terms in the Poisson sum are suppressed
by at least $1/N^{2}$ compared to the continuum operator insertion
in the loop.

The first term in the above Taylor expansion corresponds to the zero
external momentum in the loop, therefore at LO, it contributes
to the $L=0$ operator, and the sub-leading rotational invariance
breaking operators can be easily shown to be suppressed by $1/N^{2}$
using the procedure described in section \ref{sec:Scalar}. Note that
the loop integrals one needs to deal with in $g\phi^{3}$ are
more convergent than comparable integrals in $\lambda\phi^{4}$ theory, 
which 
simplifies
the discussion of the scaling of the different contributions.

The next term in the Taylor expansion of the loop integral can be
expanded in large $N$ since the integral is convergent.
The numerator has an expansion of the form
\begin{eqnarray}
{\rm Num.} & \sim & 
4N^{2}\sum_{i=1}^{3}\frac{1}{2}\sin\left(\frac{q_{i}}{N}\right)\sin\left(\frac{P_{i}}{N\Lambda}\right)
+4N^{2}\sum_{i=1}^{3}\cos\left(\frac{q_{i}}{N}\right)\sin^{2}\left(\frac{P_{i}}{2N\Lambda}\right)
\nonumber\\
&&\ =\ 
\frac{2\mathbf{P}\cdot\mathbf{q}}{\Lambda}+\frac{\left|\mathbf{P}\right|^{2}}{\Lambda^{2}}
+\mathcal{O}\left(\frac{1}{N^{2}}\right)
\ \ \ \ ,
\label{eq:68}
\end{eqnarray}
where the rotational invariance breaking terms are suppressed
by at least $1/N^{2}$, and the leading contribution to the above
integral modifies the $L=1$ matrix element, while the $L=0$ term is suppressed
by $1/\Lambda$ compared to the $L=1$ contribution. The next terms
in the Taylor expansion give rise to contributions to
the $L=2,3,...$ matrix elements at the LO 
in $1/\Lambda$, 
while the rotational invariance violating terms remain
suppressed by at least $1/N^{2}$ compared to the LO  contributions.

\section{Links on the Grid}
\label{app:linksongrid}

In this appendix, the method to evaluate the link at $\mathcal{O}\left(g\right)$
on a three-dimensional grid is outlined through an example, and the
result is generalized to other similar cases. The link is constructed 
to be the closest link to the continuum diagonal link in the continuum.

Suppose that the link lies between points $x$ and $x+\mathbf{n}a$
on a cubic lattice where: $\mathbf{n}a=a_{0}\left(Q,1,0\right)$.
$Q$ is an arbitrary integer and $a_{0}$ is a finite number denoting
the original lattice spacing which is not necessarily small. Then
the paths which make minimal area with the diagonal path can be formed
easily. Among those, the paths which are symmetric under reflection
around the midpoint of the path are desired since they have somewhat simple 
forms. One such a path in shown in fig.~\ref{fig:links}a for $Q=2$, where it is
straightforward to show that:
\begin{equation}
U_{\left(Q,1,0\right)}^{\left(1g\right)}\left(q\right)=iga_{0}e^{i\mathbf{q}\cdot\mathbf{n}a/2}
\left[A_{y}\left(q\right)+2A_{x}\left(q\right)\frac{\sin\left(Qq_{x}a_{0}/4\right)}{
\sin\left(q_{x}a_{0}/2\right)}\cos\left(\frac{Qq_{x}a_{0}}{4}+\frac{q_{y}a_{0}}{2}\right)\right]
\ \ \ .
\label{eq:69}
\end{equation}
If the lattice spacing is halved, the closest link to the diagonal
path can be obtained by adding up two paths each of the form above
with an appropriate phase factor and where $\mathbf{n}a$ is replaced
by $\mathbf{n}a/2$, fig.~\ref{fig:links}b,
\begin{eqnarray}
U_{\left(Q,1,0\right)}^{\left(1g\right)}\left(q\right)
& = & 
ig\frac{a_{0}}{2}e^{i\mathbf{q}\cdot\mathbf{n}a/2}\frac{\sin\left(\frac{\mathbf{q}\cdot\mathbf{n}a}{2}\right)}{
\sin\left(\frac{\mathbf{q}\cdot\mathbf{n}a}{4}\right)}
\left[A_{y}\left(q\right)
\phantom{\frac{\sin\left(Qq_{x}a_{0}/8\right)}{\sin\left(q_{x}a_{0}/4\right)} }
\right.\nonumber\\
&& \left.\qquad\qquad
\ +\ 
2A_{x}\left(q\right)
\frac{\sin\left(Qq_{x}a_{0}/8\right)}{\sin\left(q_{x}a_{0}/4\right)}\cos\left(\frac{Qq_{x}a_{0}}{8}
+\frac{q_{y}a_{0}}{4}\right)\right]
\ \ \ .
\label{eq:70}
\end{eqnarray}
This process can be repeated  to build extended gauge links on finer grids. 
For the general
case, where the original lattice spacing is divided by $2^{K}$, it
is not hard to show that
\begin{eqnarray}
U_{2^{K}\left(Q,1,0\right)}^{\left(1g\right)}\left(q\right)
& = & 
ig\frac{a_{0}}{2^{K}}e^{i\mathbf{q}\cdot\mathbf{n}a/2}\frac{\sin\left(\frac{\mathbf{q}\cdot\mathbf{n}a}{2}\right)}{
\sin\left(\frac{\mathbf{q}.\mathbf{n}a}{2^{K+1}}\right)}
\left[A_{y}\left(q\right)
\phantom{\frac{\sin\left(Qq_{x}a_{0}/2^{K+2}\right)}{\sin\left(q_{x}a_{0}/2^{K+1}\right)}}
\right. \nonumber\\
&& \left. \qquad\qquad
+\ 2 A_{x}\left(q\right)\frac{\sin\left(Qq_{x}a_{0}/2^{K+2}\right)}{\sin\left(q_{x}a_{0}/2^{K+1}\right)}
\cos\left(\frac{Qq_{x}a_{0}}{2^{K+2}}+\frac{q_{y}a_{0}}{2^{K+1}}\right)\right]
\ \ \ .
\label{eq:71}
\end{eqnarray}
The continuum limit is obtained by taking $K\rightarrow\infty$,
which corresponds to $a=a_{0}/2^{K}\rightarrow0$, recovering eq.~(\ref{eq:40}).
Note that after interchanging the gauge field indices properly, this
expression is applicable to a class of $\mathbf{n}$ vectors with
one zero component and $n_{i}/n_{j}=Q$ for the ratio of the remaining
components.

The above expression for the gauge link in eq.~(\ref{eq:71}) can be
generalized easily to another class of $\mathbf{n}$ vectors with
one component being equal to $Q$ and the other two components each
being one. For example for $\mathbf{n}a=a_{0}\left(Q,1,1\right)$
one obtains
\begin{eqnarray}
U_{2^{K}\left(Q,1,1\right)}^{\left(1g\right)}\left(q\right)
& = & 
ig\frac{a_{0}}{2^{K}}e^{i\mathbf{q}\cdot\mathbf{n}a/2}\frac{\sin\left(\frac{\mathbf{q}\cdot\mathbf{n}a}{2}\right)}{
\sin\left(\frac{\mathbf{q}\cdot\mathbf{n}a}{2^{K+1}}\right)}
\left[A_{z}\left(q\right)e^{iq_{y}a_{0}/2^{K+1}}+A_{y}\left(q\right)e^{-iq_{z}a_{0}/2^{K+1}}
\right.
\nonumber\\
&& \left.
+2A_{x}\left(q\right)\frac{\sin\left(Qq_{x}a_{0}/2^{K+2}\right)}{\sin\left(q_{x}a_{0}/2^{K+1}\right)}
\cos\left(\frac{Qq_{x}a_{0}}{2^{K+2}}+\frac{q_{y}a_{0}}{2^{K+1}}+\frac{q_{z}a_{0}}{2^{K+1}}\right)\right]
\ \ \ .
\label{eq:72}
\end{eqnarray}
However, since the vector $\mathbf{n}a$ is symmetric in its $y$
and $z$ components, the link has to respect this symmetry as well. 
In
fact, there exist an equivalent path which arises from
the first path by interchanging the steps in the $y$ direction and
the $z$ direction. 
Taking an average of  these two paths gives
a link which is symmetric in the $y$ and $z$ components,
\begin{eqnarray}
\bar{U}_{2^{K}\left(Q,1,1\right)}^{\left(1g\right)}\left(q\right)
& = & ig\frac{a_{0}}{2^{K}}e^{i\mathbf{q}\cdot\mathbf{n}a/2}\frac{
\sin\left(\frac{\mathbf{q}\cdot\mathbf{n}a}{2}\right)}{\sin\left(\frac{\mathbf{q}\cdot\mathbf{n}a}{2^{K+1}}\right)}
\left[A_{z}\left(q\right)\cos\left(\frac{q_{y}a_{0}}{2^{K+1}}\right)
+A_{y}\left(q\right)\cos\left(\frac{q_{z}a_{0}}{2^{K+1}}\right)
\right.
\nonumber\\
&& \left.
+2A_{x}\left(q\right)\frac{\sin\left(Qq_{x}a_{0}/2^{K+2}\right)}{
\sin\left(q_{x}a_{0}/2^{K+1}\right)}\cos\left(\frac{Qq_{x}a_{0}}{2^{K+2}}+\frac{q_{y}a_{0}}{2^{K+1}}
+\frac{q_{z}a_{0}}{2^{K+1}}\right)\right]
\ \ \ .
\label{eq:73}
\end{eqnarray}
Taking the $K\rightarrow\infty$ limit of the above link gives rise
to the rotational invariant link as well as non-continuum 
corrections which start at $\mathcal{O}\left(a^{2}\right)$.

Another class of $\mathbf{n}$ vectors are those where two components
are equal to $Q$ while the other one is equal one. For example for
$\mathbf{n}a=a_{0}\left(Q,Q,1\right)$ the link which is symmetric
with respect to $x$ and $y$ can be shown to have the form:
\begin{eqnarray}
&&\bar{U}_{2^{K}\left(Q\times Q\times1\right)}^{\left(1g\right)}
\ =\ 
ig\frac{a_{0}}{2^{K}}e^{i\mathbf{q}\cdot\mathbf{\Delta x}/2}
\frac{\sin\left(\frac{\mathbf{q}\cdot\mathbf{\Delta x}}{2}\right)}{
\sin\left(\frac{\mathbf{q}\cdot\mathbf{\Delta x}}{2^{K+1}}\right)}\times
\nonumber\\
&&
\left[2\left(A_{x}\left(q\right)
\cos\left(\frac{q_{y}a_{0}}{2^{K+1}}\right)
+A_{y}\left(q\right)
\cos\left(\frac{q_{x}a_{0}}{2^{K+1}}\right)\right)
\frac{\sin\left(
    Q(q_{x}a_{0}+q_{y}a_{0})/2^{K+2}\right)}{\sin\left((q_{x}a_{0}+q_{y}a_{0})/2^{K+1}\right)}
\right.\nonumber\\
&&\left.\qquad\qquad
\times
\cos\left(\frac{Qq_{x}a_{0}}{2^{K+2}}+\frac{Qq_{y}a_{0}}{2^{K+2}}+\frac{q_{z}a_{0}}{2^{K+1}}\right)+A_{z}\left(q\right)\right]
\ \ \ ,
\label{eq:74}
\end{eqnarray}
where $\bar{U}$ the  average of two links which are identical
upon interchanging the $x$ and $y$ coordinate axes. This link recovers
the rotational invariant link up to corrections of ${\cal O}(a^{2})$.

For $\mathbf{n}$ vectors with equal components, $\mathbf{n}a=a_{0}\left(Q,Q,Q\right)$,
there are six equivalent links which are averaged over to obtain
\begin{eqnarray}
&&
\bar{U}_{2^{K}\left(Q\times Q\times Q\right)}^{\left(1g\right)}
\ = \ 
ig\frac{a_{0}}{2^{K}}e^{i\mathbf{q}\cdot\mathbf{\Delta
    x}/2}\frac{\sin\left(\frac{\mathbf{q}\cdot\mathbf{\Delta x}}{2}\right)}
{\sin\left(\frac{\mathbf{q}\cdot\mathbf{\Delta x}}{2^{K+1}}\right)}\times
\nonumber\\
&&
\left[2\left(A_{x}\left(q\right)\cos\left(\frac{q_{y}a_{0}}{2^{K+1}}+\frac{q_{z}a_{0}}{2^{K+1}}\right)+A_{y}\left(q\right)
\cos\left(\frac{q_{x}a_{0}}{2^{K+1}}+\frac{q_{z}a_{0}}{2^{K+1}}\right)
+A_{z}\left(q\right)\cos\left(\frac{q_{x}a_{0}}{2^{K+1}}+\frac{q_{y}a_{0}}{2^{K+1}}\right)\right)\right.
\nonumber\\
&&
\times\left.\frac{
\sin\left((Qq_{x}a_{0}+Qq_{y}a_{0}+Qq_{z}a_{0})/2^{K+2}\right)}
{\sin\left((q_{x}a_{0}+q_{y}a_{0}+q_{z}a_{0})/2^{K+1}\right)}
\cos\left(\frac{Qq_{x}a_{0}}{2^{K+2}}+\frac{Qq_{y}a_{0}}{2^{K+2}}+\frac{Qq_{z}a_{0}}{2^{K+2}}
\right)
\right]
\ ,
\label{eq:75}
\end{eqnarray}
which results in $\mathcal{O}\left(a^{2}\right)$ corrections to the
rotational invariant continuum path.
It is the case that determining
the link for a general extended path  is  quite involved, but the
general trend that the  deviation  from the
rotationally invariant continuum path is  ${\cal O}(a^{2})$
is anticipated.

\bibliography{bibi}
\end{document}